\documentclass[aps,prb, reprint,superscriptaddress]{revtex4-2}

\renewcommand{\selectlanguage}[1]{}

\usepackage{upgreek,mathtools}
\usepackage{color}
\usepackage{graphicx}
\usepackage{nccmath}
\usepackage{bm}
\usepackage{hyperref}
\hypersetup{colorlinks,breaklinks,
            urlcolor=[rgb]{0,0,0.36},
            linkcolor=[rgb]{0,0,0.36},
            citecolor=[rgb]{0,0,0.36}}

\usepackage[mathlines]{lineno}
\usepackage{dcolumn}
\usepackage{mathtools}  
\usepackage{dsfont}
\usepackage{comment}
\usepackage{physics}
\usepackage{xcolor}
\usepackage{adjustbox}

\newcommand{\varphicl}{\varphi^{\rm cl}}
\newcommand{\varphiq}{\varphi^{\rm q}}

\newcommand{\Jcl}{J^{\rm cl}}
\newcommand{\Jq}{J^{\rm q}}

\newcommand{\Harvard}{Department of Physics, Harvard University, Cambridge, Massachusetts 02138, USA.}

\newcommand{\MaxP}{Max Planck Institute for the Structure and Dynamics of Matter, Luruper Chaussee 149, 22761 Hamburg, Germany.}

\newcommand{\BNL}{Condensed Matter Physics and Materials Science Division, Brookhaven National Laboratory, Upton, New York 11973, USA.}

\newcommand{\Oxford}{Clarendon Laboratory, University of Oxford, Parks Road, Oxford OX1 3PU, UK.}

\newcommand{\ETH}{Institute for Theoretical Physics, ETH Zurich, 8093 Zurich, Switzerland.}

\begin{document}

\title{Two-dimensional spectroscopy of bosonic collective excitations in disordered many-body systems}

\author{Alex~G\'{o}mez~Salvador}
\thanks{These authors contributed equally to this work.}
\affiliation{\ETH}

\author{Ivan~Morera}
\thanks{These authors contributed equally to this work.}
\affiliation{\ETH}

\author{Marios~H.~Michael}

\affiliation{\MaxP}

\author{Pavel~E.~Dolgirev}

\affiliation{\Harvard}

\author{Danica~Pavicevic}
\affiliation{\MaxP}

\author{Albert~Liu}

\affiliation{\BNL}


\author{Andrea~Cavalleri}
\affiliation{\MaxP}
\affiliation{\Oxford}

\author{Eugene~Demler}
\affiliation{\ETH}
\date{\today}

\begin{abstract}
    {We present a novel theoretical approach for computing and analyzing two-dimensional spectroscopy of bosonic collective excitations in disordered many-body systems. Specifically, we employ the Keldysh formalism to derive, within a non-pertubative treatment of disorder effects, the third-order nonlinear response and obtain two-dimensional spectroscopy maps. In the weak nonlinear regime of our formalism, we demonstrate the ability of the echo peak to distinguish between elastic and inelastic scattering processes, in perfect agreement with the intuition developed in isolated two-level systems.
    Furthermore, we discuss unique many-body effects on the echo peak signature arising from interaction induced quantum fluctuations. In particular, we show that these quantum fluctuations induce a finite nonrephasable broadening and examine how the echo peak is influenced by the attractive or repulsive nature of the collective excitations.}
    \end{abstract}

\maketitle
\section{Introduction}
Linear response theory underlies most of the traditional experimental techniques in quantum many-body physics. In such experiments, results can be interpreted from the perspective of two-point correlation functions of appropriate operators. In the case of electron systems, paradigmatic examples include density operators for X-ray scattering~\cite{guinier1994_xray,Comin2016}, current operators for optical spectroscopy~\cite{Jimenez2016} and transport~\cite{robson2017fundamentals,Pekola_heat_transport_2021}, and electron creation and annihilation operators for STM~\cite{Binning_1987_review_STM} and ARPES~\cite{Damascelli2003} experiments. Collective excitations  manifest themselves as peaks in the response functions and provide clear signatures of underlying many-body states. Investigations of collective modes using linear response probes have been ubiquitous in physics, chemistry, biology, and material science.
However, despite their immense accomplishments, these techniques also present certain shortcomings. For instance, understanding the origin of the excitation broadening can be challenging since different mechanisms often result in similar lineshapes. The extension of linear response measurements to the nonlinear regime, and the consequent access to higher-order correlators, has opened the door to circumvent some of these limitations. Of special relevance are the accomplishments of nuclear magnetic resonance (NMR)~\cite{ernst1990principles,friebolin1991basic,book_Zerbe,Reif2021,book_McRobbie,Kim2010,keeler2010understandingNMR} and its optical analogs termed multidimensional coherent spectroscopies (MDCS)~\cite{Mukamel,hamm_zanni_2011,Cundiff2013,Bruder2019,Mukamel2009,Petti2018,Cundiff2012,fresch2023review}. For instance, these approaches have been employed to identify different interaction mechanisms between excitations~\cite{lomsadze2017frequency,liu2018vibrational,muir2022interactions}, resolve energy transfer pathways~\cite{singh2016quantifying,hybl1998review,engel2007evidence,Johnson2019pathways}, classify bound states~\cite{stone2009two,hao2017trion,hao2017neutral}, and disentangle homogeneous and inhomogeneous broadening~\cite{Bristow2011,moody2015intrinsic,huang2023quantum}.

In condensed matter systems, the characteristic low-energy excitations typically lie in the terahertz frequency (THz) range, including Josephson plasmons in high temperature superconductors~\cite{Marios_parametric_2020,Sellati2023GeneralizedJPs,fiore2024investigating,sellati2024optical,taherian2024squeezed,Kim2024Tracing}, magnons~\cite{zakeri2018magnons_terahertz,peedu2022terahertz_magnon,mehra2024myriad}, and cyclotron orbits in 2D electron gas~\cite{lloyd2014cyclotrob_thz,scalari2012cyclotron_thz}, among many others.  Historically, the well-known THz gap~\cite{book_Pavlidis} has hindered the use of nonlinear probes in studying said low-energy excitations. However, the recent progress in terahertz technology~\cite{Nicoletti2016,Fulop2020} has led to the development of a terahertz analogue of multidimensional optical spectroscopies, known as two-dimensional terahertz spectroscopy (2DTS)~\cite{Lu2016Photonecho,Maag2016Cyclotron,Houver2019ballisticConduction,Mahmood2021,Pal2021Origin,Lin2022Mapping,Luo2023Tomography,Reimann2021,Soranzio2024Nonlinear,Liu2024Artifacts,Liu2025_perspective}. In recent years, this technique has been applied to a wide-range of condensed matter systems, including superconductors~\cite{Novelli_persistent_2020,Liu_2023_echo,Sijie_2023,puviani2022quench,puviani2023quench,katsumi_revealing_2024_benfatto,puviani2024theory,katsumi2024amplitudemodemultigapsuperconductor}, correlated metals~\cite{barbalas2023energy} and insulators~\cite{chen2024multidimensionalcoherentspectroscopycorrelated}, ferroics~\cite{Lu2017,Lin2022,zhang2024terahertz1,zhang2024terahertz2,parameswaran2020asymptotically1}, topological materials~\cite{Blank2023,wan2019resolving,liebman2023multiphotonspectroscopydynamicalaxion}, and spin liquids~\cite{Choi2020SpinLiquid,mcginley2022signatures,McGinley2024Anomalous} and ices~\cite{potts2024signaturesspinondynamicsphase}. 


Similarly to MDCS, 2DTS is poised to be a promising technique to disentangle different sources of broadening in a correlated many-body system.
However, the applicability of the standard interpretation in terms of homogeneous and inhomogeneous broadening in a system of isolated two-level systems (ITLS) becomes questionable when studying collective modes. 
This is particularly relevant when examining the impact of random spatial inhomogeneities on collective excitations.

This paper provides a new theoretical framework for analyzing 2DTS for collective excitations in the presence of static spatial disorder. 
Employing a self-consistent field-theoretical approach in the Keldysh formalism we compute the third-order nonlinear response of a many-body system in the presence of disorder and obtain the characteristic asymmetric (almond-like) echo peak signature in the 2DTS protocol.
In particular, we demonstrate the non-perturbative nature of the rephasing physics and the necessity to consider the infinite series of non-crossing disorder diagrams.
In the weak nonlinear regime, the self-consistent nature of our framework allows us to derive a general analytical relation between the disorder dressed vertex and the self-energies associated with elastic and inelastic scatterings, Eq.~\eqref{eqn: vertex_self_energy_classical}. This expression recovers the well-known echo peak phenomenology from ITLS ensembles, which emerges from the interplay between the two broadening mechanisms and explains the characteristic almond-like signature in the presence of static disorder. Furthermore, we generalize the previous expression to account for quantum fluctuations arising from many-body correlations, Eq~\eqref{eqn: vertex_self_energy_quantum}. We discuss how strong interaction induced quantum fluctuations affect the echo peak phenomenology, and in particular act as a source of broadening which cannot be rephased, rendering perfect rephasing fundamentally unattainable in interacting many-body systems. 
Finally, we discuss how the attractive or repulsive nature of the collective excitations modify the echo peak signature. 


The rest of the paper is organized as follows. In Sec.~\ref{Sec: intuition}, we introduce the 2DTS protocol and provide an intuitive discussion of the signatures of disorder in the 2D maps. In Sec.~\ref{sec: keldysh}, we describe our theoretical approach to compute the nonlinear 2D maps within the Keldysh path integral formalism and provide a general equation for the third-order susceptibility in terms of the linear susceptibility and the ``dressed" vertex. In Sec.~\ref{sec: RPA}, we outline a conserving calculation of the third-order susceptibility and the associated 2D map for a system without disorder. Then, in Sec.~\ref{sec: elastic}, we detail the calculations of the 2D map considering static disorder. In Sec.~\ref{sec: classical_limit}, we discuss the signatures of the 2D map in the weak nonlinear regime, and in Sec.~\ref{sec: quantum_limit}, the effect of interaction induced quantum fluctuations. Finally, in Sec.~\ref{sec: conclusions}, we present our conclusions and outlook.
\begin{figure}
    \centering
    \includegraphics[width=1\linewidth]{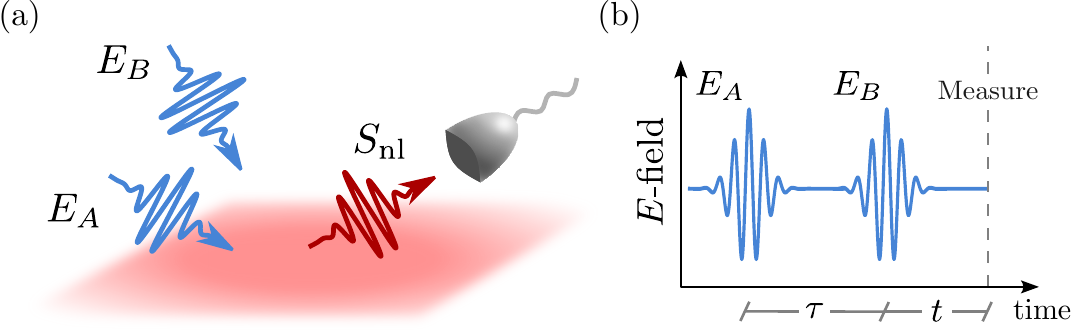}
    \caption{(a) Cartoon of the standard two-dimensional terahertz spectroscopy protocol. External perturbations $E_A$ and $E_B$ are sent with a relative time delay $\tau$ (b). After a waiting $t$ with respect to the arrival of the second pulse, the measurement of the resulting nonlinear signal $S_{\rm nl}$ is performed.}
    \label{fig:intro_2d}
\end{figure}
\section{Echo peak in many-body systems: ITLS and collective excitations}\label{Sec: intuition}
To set notations for subsequent discussion of disordered quantum many-body systems, we first describe the standard 2DTS protocol and introduce the two-dimensional (2D) map. Then, we provide a qualitative discussion of the difference in the signatures of disorder in the 2D map for ITLS and collective excitations, and motivate the necessity for a reformulation of the interpretation of the disorder effects.

The standard 2DTS protocol can be summarized as follows: An external perturbation consisting of two identical excitation pulses, denoted $E_A$ and $E_B$, separated by a time delay $\tau>0$, is sent towards the sample. After the arrival of the second pulse, the system is left to evolve unperturbed for time $t>0$, after which the measurement is performed to obtain the signal $S(\tau,t)$, see Fig.~\ref{fig:intro_2d}. In this type of protocols, both the linear contributions in $E_A$ or $E_B$, and the self-nonlinearities proportional to $E_A^3$ and $E_B^3$ are usually filtered out experimentally. This leaves behind a purely nonlinear signal $S_{\rm nl}(\tau,t)$ with terms proportional to $E_A^2 E_B$ $(A)$ and $E_A E_B^2$ $(B)$ . Assuming the pulses to be perfect Dirac delta functions, the $E_A^2 E_B$ signal can be obtained from the third-order nonlinear response $\chi^{(3)}$ via
\begin{equation}
    S_{\rm nl}^{A}(t,\tau) =3\int\prod_{i=1}^3\frac{d\omega_i}{2\pi}\,\chi^{(3)}(\omega_3,\omega_2,\omega_1)e^{-i(\omega_1+\omega_2)\tau}e^{-i\bar{\omega}t},
    \label{eqn: S_t_tau_A}
\end{equation}
and similarly for the $E_AE_B^2$ signal:
\begin{equation}
    S_{\rm nl}^{B}(t,\tau) =3\int\prod_{i=1}^3\frac{d\omega_i}{2\pi}\,\chi^{(3)}(\omega_3,\omega_2,\omega_1)e^{-i\omega_1\tau}e^{-i\bar{\omega}t},
    \label{eqn: S_t_tau_B}
\end{equation}
where $\bar{\omega}=\omega_1+\omega_2+\omega_3$. The 2D map is subsequently obtained upon performing a double Fourier transform with respect to $t$ and $\tau$ restricted to $t>0$ and $\tau>0$. Of special interest in multidimensional spectroscopy is the so-called echo peak, or rephasing peak, a nonlinear signal contained in the $E_AE_B^2$ contribution of the 2D map, analogous in nature to the spin echo~\cite{Hahn_1950}. This peak has been successfully employed in NMR and MDCS to disentangle and quantify the presence and strength of homogeneous and inhomogeneous broadening~\cite{Siemens_2010}. We now discuss and contrast the echo peaks for ITLS and collective excitations

\begin{figure}
    \centering
    \includegraphics[width=1\linewidth]{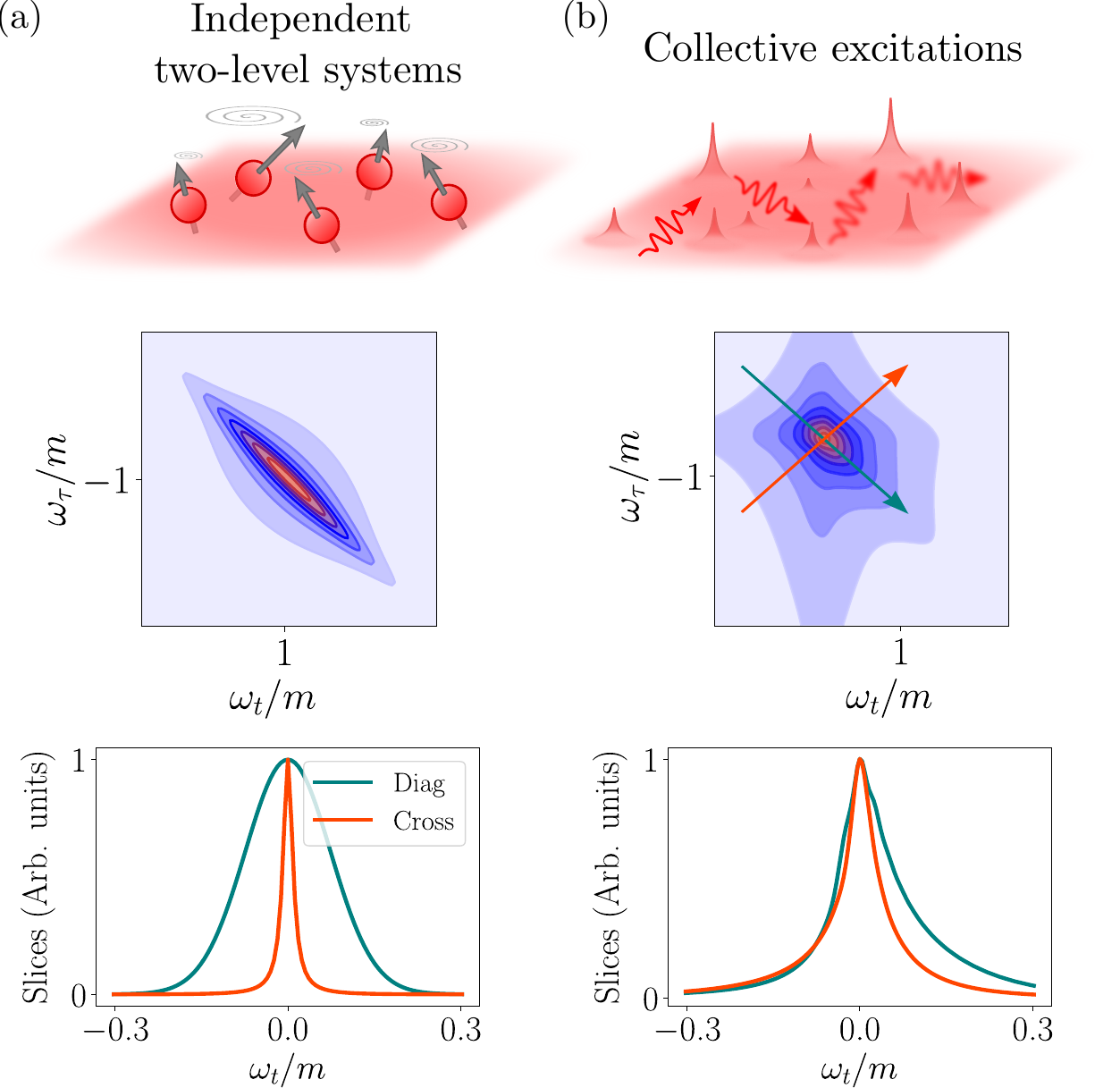}
    \caption{(a)-(b) Cartoon of the effects of disorder on ITLS and collective excitations. In (a) an ensemble of two-level systems (spin precession) have a Larmor frequency that depends on their position in space. Their collective signal exhibits broadening due to decoherence, i.e. oscillations of different ITLS in the ensemble occurring at different  frequencies. In (b) a collective excitation undergoes elastic scattering events due to the presence of disorder in the sample. This series of scattering events results in a finite lifetime for the excitation. Below, the echo peak and slices along the diagonal (teal) and cross-diagonal (orange) of the echo peak for ITLS and collective excitations ($V_0^2/m^2v^2=2$, $g/m^2=0.25$, and $\gamma/m=0.015$). When all scattering events come from static disorder, the echo peak for ITLS is strongly asymmetric with a cut along the cross-diagonal becoming infinitely sharp; for collective excitations the asymmetry is less pronounced and the cut along the cross-diagonal presents a finite width due to the presence of many-body correlations. Note that for the sake of visualisation, we have included a finite small damping to the ITLS echo peak of $\gamma/m=0.015$, since otherwise, it would be infinitely sharp along the cross diagonal. }
    \label{fig: intro_echoes}
\end{figure}

First, consider an ensemble of non-interacting ITLS spatially distributed such that their Larmor frequencies depend on their position, see Fig.~\ref{fig: intro_echoes}(a). 
In the simplest theoretical treatment, one introduces the linewidth $\gamma$, characterizing the $T_1$ and $T_2$ processes (homogeneous broadening), and assumes a Gaussian distribution for the Larmor frequencies with mean $\omega_0$ and variance $\sigma$ (inhomogeneous broadening). 
Under these assumptions, the time-domain echo nonlinearity in a 2-pulse protocol can be obtained exactly~\cite{Siemens_2010}:
\begin{equation}
    S(\tau,t) \sim \theta(\tau)\theta(t) e^{-i\omega_0(t-\tau)}e^{-\gamma(t+\tau)}e^{-\sigma^2(t-\tau)^2/2}.
    \label{eqn: echo_ITLS}
\end{equation}
The success of two-dimensional spectroscopy in disentangling homogeneous and inhomogenoeus broadening is due to the rephasing nature of the echo peak, which exhibits a distinct time-dependence on the terms proportional to $\gamma$ and $\sigma$. This distinct time-dependence results not only in a characteristic almond-like shape for the echo peak, but also enables a simultaneous fitting procedure for $\gamma$ and $\sigma$ based on two orthogonal diagonal cuts of the echo peak along $\omega_t\pm\omega_\tau$~\cite{Siemens_2010}, see Fig.~\ref{fig: intro_echoes}(a) and (b).

Consider now a many-body system in the absence of disorder where collective modes are infinitely long lived. When the system is driven by light, excitations are created or annihilated with zero momentum, causing them to oscillate at the mass (resonance) energy without any damping. However, if the medium is disordered, for example, due to the presence of static impurities, elastic scattering events will result in a finite excitation lifetime. It is useful to contrast collective mode eigenstates in the presence of static disorder with propagators at specific momenta. While the former have well-defined energies and infinite lifetimes (in the absence of interactions between modes), the latter have finite broadening, corresponding to finite lifetimes, and described mathematically by the self-energies of individual excitations $\Sigma$. In contrast to the standard ITLS scenario, the intrinsic interactions of nonlinear excitations generate many-body correlations between modes, resulting in additional broadening  even in the absence of homogeneous damping. This invalidates the usual distinction between homogeneous and inhomogeneous broadening, and therefore an interpretation based on such a separation can lead to misleading conclusions. As we show in this work, distinguishing between elastic and inelastic scattering events, while considering the effects of interaction induced quantum fluctuations, provides a more suitable separation and interpretation of the broadening mechanisms.

Having established the breakdown of the separation between homogeneous and inhomogeneous broadening, we now address the rephasing nature of the echo peak in a many-body system. In the simplest scenario, the essence of the echo peak and its characteristic almond shape relies on the system evolving with the same frequency during time delays $\tau$ and $t$, c.f.~\eqref{eqn: echo_ITLS}. If this is to originate from scattering events, then these scatterings must correlate the excitations present during $\tau$ and $t$, causing them to explore states with the same energy. 
Intuitively, two distinct excitations are created by pulses A and B, respectively. The echo response arises from correlations between the scattering processes of these two excitations—for example, when both scatter off the same impurities. Consequently, it is essential to analyze the vertex corrections to the interaction $\Gamma$ due to the presence of disorder.
However, not all vertex corrections have a rephasing nature, implying that either only a partial rephasing is possible or the rephasing diagrams are strongly peaked at the echo diagonal to compensate for the finite lifetime induced by disorder. As we discuss in Sec.~\ref{sec: classical_limit}, in the absence of inelastic scattering and neglecting interaction induced quantum fluctuations, the rephasing contributions indeed present a divergence in the echo sector, thereby enabling perfect rephasing in this limiting case. However, when such quantum fluctuations are taken into account, we demonstrate that perfect rephasing is fundamentally unattainable, see Sec~\ref{sec: quantum_limit}.
In conclusion, rephasing in collective excitations is possible, although only partially, and thus one can also expect a similar, but not identical, almond-like signature in the echo peak, compare Fig.~\ref{fig: intro_echoes}(a) and (b). Furthermore, from a theoretical point of view, disorder induced vertex corrections must be taken into account to capture rephasing effects.

\section{Nonlinear response in the Keldysh formalism}\label{sec: keldysh}
This section provides a general introduction to the computation of third-order nonlinear responses in the Keldysh formalism, alongside an intuitive understanding of the diagrammatics involved. We follow closely throughout~\cite{dupuis2023field,kamenev2023field,peskin2018introduction,berges2004introduction} for the field theory discussion and~\cite{Mukamel} for the nonlinear response. The structure of this section is as follows: In subsection~\ref{subsec: general_nonlinear}, we introduce the generic $\varphi^4$ model to study the nonlinear response of collective bosonic excitations and discuss how to express nonlinear response functions as correlators in the Keldysh formalism. In~\ref{subsec: diagramm_chi_3}, inspired by the lowest order approximation in interaction strength to the nonlinear response, we argue for the general diagrammatic structure of $\chi^{(3)}$. 

\subsection{Model and General considerations}\label{subsec: general_nonlinear}
We begin by considering a non-interacting relativistic action for a real scalar bosonic field $\varphi$ of the form
\begin{equation}
    S_0[\varphi] = \int_\mathcal{C}dt\int d^d \boldsymbol{r}\, \frac{1}{2}\left[ \left(\partial_t\varphi\right)^2-v^2\left( \nabla\varphi \right)^2 - m_0^2\varphi^2 \right],
\end{equation}
where $\mathcal{C}$ denotes the Keldysh contour, $v$ the velocity, and $m_0$ the bare mass.
To perform the integral along the Keldysh contour we introduce two bosonic fields $\varphi^+$ and $\varphi^-$ which reside in the forward and backward parts of the time contour, respectively. The non-interacting retarded, advanced and Keldysh Green's functions are given in the Keldysh rotated basis $\varphi^{\pm}=\varphi^{\rm cl}\pm\varphi^{\rm q}$ by
\begin{equation}
    i\mathcal{D}_0^{\alpha\beta}(\boldsymbol{r}-\boldsymbol{r}',t-t')=\int \boldsymbol{D}[\varphi] \, \varphi^{\alpha}(\boldsymbol{r},t) \varphi^{\beta}(\boldsymbol{r}',t) e^{iS_0[\varphicl,\varphiq]},
\end{equation}
with $\alpha,\beta=\{\rm cl,\rm q\}$, which further presents causality structure typical of the Keldysh propagator:
\begin{equation}
    \mathcal{D}_0^{\alpha\beta} = \begin{pmatrix} \mathcal{D}_0^{\rm K}(\boldsymbol{r}-\boldsymbol{r}',t-t') & \mathcal{D}_0^{\rm R}(\boldsymbol{r}-\boldsymbol{r}',t-t') \\ \mathcal{D}_0^{\rm A}(\boldsymbol{r}-\boldsymbol{r}',t-t') & 0 \end{pmatrix}.
\end{equation}
Note that $\boldsymbol{D}$ includes $1/\Tr{\rho_0}$, with $\rho_0$ the equilibrium density matrix, following the standard Keldysh path integral measure~\cite{kamenev2023field}. Equivalently, the Green's functions in frequency-momentum representation are given by
\begin{gather}
    \mathcal{D}^{R/A}_0(k,\omega) = \frac{1}{2}\frac{1}{(\omega\pm i\eta)^2-\epsilon_{\boldsymbol{k}}^2}, \notag \\ \mathcal{D}_0^K(k,\omega) = F(\omega)\left[\mathcal{D}^R_0(k,\omega)-\mathcal{D}^A_0(k,\omega)\right]
    \notag \\=-\frac{i\pi}{2\epsilon_{\boldsymbol{k}}}\coth(\frac{\omega}{2T})\left[\delta(\omega-\epsilon_{\boldsymbol{k}})-\delta(\omega+\epsilon_{\boldsymbol{k}})\right].
\end{gather}
Here, $\epsilon_{\boldsymbol{k}} = \sqrt{m_0^2+v^2k^2}$ is the bare energy-momentum dispersion relation, $\eta=0^+$ is the (anti-)causal regularization, $F(\omega) = \coth(\omega/2T)$ is the equilibrium distribution function, and $T$ is the temperature of the system. We consider an interaction between the bosons given by a generic $\varphi^4$ term:
\begin{align}
    S_{\rm int}[\varphi] &= -\int_{\mathcal{C}}dt\int d^d \boldsymbol{r}\, \frac{g}{4!}\varphi^4 \notag\\ &= -\int_{-\infty}^{\infty} dt\int d^d \boldsymbol{r} \, \frac{g}{3}\left[ \varphi^{cl}(\varphi^q)^3 + (\varphi^{cl})^3\varphi^q \right].
\end{align}
In order to evaluate the response of the system to an external perturbation, we further introduce a drive/source term in the action given by
\begin{align}
    S_{\rm drive}[\varphi,j]&= \int_{\mathcal{C}}dt\int d^d \boldsymbol{r}\, j\varphi \notag \\ &= \int_{-\infty}^{\infty} dt\int d^d \boldsymbol{r}\, 2\left( j^{cl}\varphi^q + j^q\varphi^{cl} \right),
\end{align}
and define the partition function $Z[j]$ as
\begin{equation}
    Z[j] = \int \boldsymbol{D}[\varphicl,\varphiq]e^{iS[\varphicl,\varphiq,j^{\rm cl},j^{\rm q}]+iS_{\rm drive}[\varphi,j]},
\end{equation}
where $S[\varphi,j] = S_0[\varphi] +S_{\rm int}[\varphi]$. In the Keldysh formalism, the measured physical field is given by $\expval{\varphicl(t)}$ and the response functions are obtained by differentiation with respect to the physical sources $j^{\rm cl}$. The third-order response function in frequency-momentum representation can be obtained by differentiating with respect to classical sources 
\begin{widetext}
\begin{align}
    \chi_{\boldsymbol{k};\boldsymbol{k}_1,\boldsymbol{k}_2,\boldsymbol{k}_3}^{(3)}(\omega;\omega_3,\omega_2,\omega_1) &= \frac{\delta^{3}\expval{\varphi^{cl}_{\boldsymbol{k}}(\omega)}}{\delta j^{cl}_{\boldsymbol{k}_3}(\omega_3)\delta j^{cl}_{\boldsymbol{k}_2}(\omega_2)\delta j^{cl}_{\boldsymbol{k}_1}(\omega_1)}=-\frac{i}{2}\frac{\delta^4Z}{\delta j^{q}_{\boldsymbol{k}}(\omega)\delta j_{\boldsymbol{k}_3}^{cl}(\omega_3)\delta j^{cl}_{\boldsymbol{k}_2}(\omega_2)\delta j^{cl}_{\boldsymbol{k}_1}(\omega_1)}\eval_{j=0} \notag \\ &= -8i\expval{\varphi^{cl}_{\boldsymbol{k}}(\omega)\varphi^{q}_{\boldsymbol{k}_3}(\omega_3)\varphi^{q}_{\boldsymbol{k}_2}(\omega_2)\varphi^{q}_{\boldsymbol{k}_1}(\omega_1)},
    \label{eqn: chi3_greens}
\end{align}
such that the expectation value of the boson field at time $t$ is given by
\begin{align}
    \expval{\varphicl_{\boldsymbol{k}}(t)}^{(3)} & =\frac{1}{L^{3d}}\sum_{\boldsymbol{k}_1,\boldsymbol{k}_2,\boldsymbol{k}_3}\int\frac{d\omega_1}{2\pi} \int\frac{d\omega_2}{2\pi} \int\frac{d\omega_3}{2\pi} \, \chi^{(3)}_{\boldsymbol{k};\boldsymbol{k}_1,\boldsymbol{k}_2,\boldsymbol{k}_3}(\omega;\omega_3,\omega_2,\omega_1)\, j_{\boldsymbol{k}_3}(\omega_3) j_{\boldsymbol{k}_2}(\omega_2) j_{\boldsymbol{k}_1}(\omega_1) \,e^{-i\omega t}.
    \label{eqn: phi3}
\end{align}
\end{widetext}
For time- and space-translationally invariant systems $\expval{\varphi^{cl}_{\boldsymbol{k}}(\omega)\varphi^{q}_{\boldsymbol{k}_3}(\omega_3)\varphi^{q}_{\boldsymbol{k}_2}(\omega_2)\varphi^{q}_{\boldsymbol{k}_1}(\omega_1)}\sim\delta(\omega-\bar{\omega})\delta^d(\boldsymbol{k}-\bar{\boldsymbol{k}})$, with $\bar{\omega}=\omega_1+\omega_2+\omega_3$ and $\bar{\boldsymbol{k}}=\boldsymbol{k}_1+\boldsymbol{k}_2+\boldsymbol{k}_3$. Note that causality imposes that $\chi^{(n)}(t_n,t_{n-1},\dots,t_1) =0 $ if $t_j<0$. Furthermore, in the absence of interactions, it is clear that Eq.~\eqref{eqn: chi3_greens} vanishes due to the presence of $\expval{\varphiq\varphiq}$ in all possible Wick contractions. We note that some of the results presented here can also be found in the context of the one-dimensional transverse field Ising model in~\cite{Spinon_Hart_2023}.

\subsection{Diagrammatic representation of \texorpdfstring{$\chi^{(3)}$}{x(3)}}\label{subsec: diagramm_chi_3}

In this work, we focus on the situation where the field perturbing the system, $j$, couples to the zero-momentum component of the bosonic field, $\varphi$, and measurements of the bosonic mode are also performed at $\boldsymbol{k}=0$. This is indeed the standard scenario when light is used to drive an excitation in a condensed matter system, for example in Ref.~\cite{Liu_2023_echo}. Therefore, we drop the spatial/momentum index in the rest of this subsection, understanding that light always couples to the collective excitation at zero momentum. 

To lowest order in the interaction, the third-order nonlinear response is given by (see Appendix~\ref{Appendix: tree_level} for a detailed derivation):
\begin{equation}
    \chi^{(3)}_{\rm mf}(\omega_1,\omega_2,\omega_3)
    =-g\,\chi^{(1)}(\omega_1)\chi^{(1)}(\omega_2)\chi^{(1)}(\omega_3)\chi^{(1)}(\bar{\omega}),
    \label{eqn: mean_field_chi_3}
\end{equation}
which readily recovers the ``mean field'' response obtained in~\cite{salvador2024principles} from an equations of motion formulation. The diagrammatic representation of $\chi^{(3)}$ in Eq.~\eqref{eqn: mean_field_chi_3} allows for a rather intuitive interpretation of the processes involved in the nonlinear response. The first-order perturbative calculation can be represented as
\begin{equation}
    \chi^{(3)}_{\rm mf}(\omega_3,\omega_2,\omega_1) = \quad \vcenter{\hbox{\includegraphics[scale=0.65]{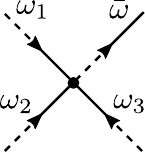}}}\quad .
    \label{eqn: mf_diag}
\end{equation}
The dashed-to-solid line represents the retarded Green's function of the boson and thus establishes the arrow of time. The dot simply denotes the nonlinearity strength $g$ coming from the $\varphi^4$ term in the action. 
Thanks to the built-in causality, the diagram Eq.~\eqref{eqn: mf_diag} can naturally be interpreted as a four-wave mixing process. First, three collective modes with frequencies $\omega_1$, $\omega_2$, and $\omega_3$ are excited in the material (note the possibility of oscillating at negative frequencies). At some point, the three excited modes are combined via a four-wave mixing process into a new mode with frequency $\bar{\omega}=\omega_1+\omega_2+\omega_1$ which then, propagates through the material; this is the mode that is eventually measured by the protocol. The separation between the incoming and outgoing modes is naturally built in the Keldysh formalism and provides an intuitive understanding of the processes involved in the nonlinear response. 

From the diagrammatic representation of $\chi^{(3)}$ in Eq.~\eqref{eqn: mean_field_chi_3} we expect that the full nonlinear susceptibility can be obtained by dressing the external propagators via the self-energy $\Sigma$ and by including vertex corrections to the four-wave mixing process $\Gamma$, see Appendix~\ref{Appendix: effective_action} for a formal proof. That is, the external leg structure of the diagram remains the same; this can be depicted diagrammatically as
\begin{align}
    \chi^{(3)}(\omega_3,\omega_2,\omega_1) &= i\Gamma(\omega_3,\omega_2,\omega_1) \prod_i^4 \chi^{(1)}(\omega_i)  \notag\\&= \quad \vcenter{\hbox{\includegraphics[scale=0.65]{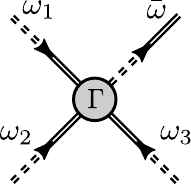}}}\quad,
    \label{eqn: general_diag}
\end{align}
with the dressed propagator given by Dyson's equation:
\begin{equation}
    \vcenter{\hbox{\includegraphics[scale=0.7]{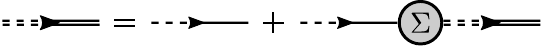}}}\,.
    \label{eqn: Dyson}
\end{equation}
We thus conclude that once $\Gamma$ and $\chi^{(1)}$ are known, we have fully determined the third-order nonlinear response.
Furthermore, from this interpretation of the $\chi^{(3)}$ response, it becomes clear that the location of the peaks in the 2D map is determined by the excitation linewidth, or equivalently by the support of the imaginary part of the self-energy. The vertex $\Gamma$ can only ``redistribute'' the weight in the vicinity of said support.

In this work, we adopt a conserving nonperturbative approach based on the Baym-Kadanoff construction~\cite{Baym_Kadanoff_1961,Baym_1962}. This allows us to obtain the vertex $\Gamma$ and linear response $\chi^{(1)}$ in a self-consistent manner. However, we envision that other nonperturbative approaches~\cite{WETTERICH199390,BERGES2002223,DUPUIS20211}, such as the functional renormalization group, could be employed to obtain the 2D map beyond our current analysis. 

To demonstrate the capabilities of our framework, we provide in the next section a derivation of $\chi^{(3)}$ in the absence of disorder, which already shows the effects of many-body correlations in the 2D map. The reader interested only in the effects of disorder in the nonlinear response, can directly proceed to Sec.~\ref{sec: elastic}.

\section{Random Phase Approximation}\label{sec: RPA}
\begin{figure}
    \centering
    \includegraphics[width=0.95\linewidth]{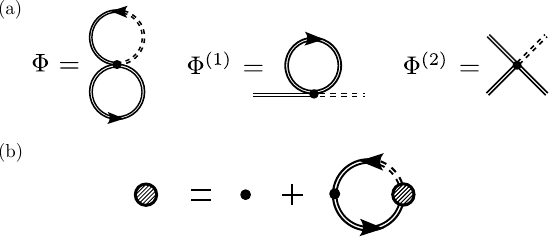}
    \caption{(a) Luttinger-Ward functional approximation to lowest order in interaction and subsequent contributions to the self-energy and vertex. (b) Diagrammatic representation of the Bethe-Salpeter equation for the full vertex within the RPA approximation.}
    \label{fig: 2PI_RPA}
\end{figure}
As an initial demonstration of the computational capabilities of the formalism developed in the previous section we consider, in the absence of static disorder, the lowest-order conserving approximation in $g$, corresponding to the Random Phase Approximation (RPA). This section is divided into two subsections. The first subsection focuses on the detailed calculation of $\chi^{(3)}$ within the RPA, providing all the relevant diagrams and deriving the corresponding self-energy and vertex corrections. The second subsection examines the signatures of the 2D map peaks through both analytical results and numerical simulations, while elucidating important connections with linear response measurements and previous theoretical work~\cite{salvador2024principles}.
\subsection{\texorpdfstring{$\chi^{(3)}$}{x(3)} within RPA}
In this subsection, we employ the Keldysh formalism to compute $\chi^{(3)}$ within the (RPA). In order to have a conserving perturbative expansion, we begin by writing the lowest order approximation of the Luttinger-Ward functional and its resulting contributions to the self-energy and vertex, see Fig.~\ref{fig: 2PI_RPA}. 
\subsubsection{Self-energy}
The self-energy calculation results in a constant term (potentially diverging depending on the dimensionality)
\begin{equation}
    \Sigma_g(\omega) = \frac{\delta\Phi[\mathcal{D}]}{\delta\mathcal{D}^R(\omega)}=\vcenter{\hbox{\includegraphics[scale=0.75]{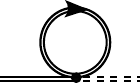}}}=\frac{ig}{L^d}\sum_{\boldsymbol{k}}\int\frac{d\Omega}{2\pi}\mathcal{D}_{\boldsymbol{k}}^K(\Omega)
    \label{eqn: self_energy_RPA}
\end{equation}
which renormalizes the mass. We subsequently fix the renormalized bare mass $m_0$ to match the physically observed mass $m$.
\subsubsection{Vertex}
The bare vertex is obtained by cutting open two Green's function from the Luttinger-Ward functional, see (a) in Fig.~\ref{fig: 2PI_RPA}, and it is simply given by
\begin{equation}
    \Gamma^{(0)}=\frac{\delta^2\Phi[\mathcal{D}]}{\delta\mathcal{D}^K(\omega)\delta\mathcal{D}^R(\omega)}=\vcenter{\hbox{\includegraphics[scale=0.7]{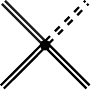}}}=ig,
    \label{eqn: bare_vertex_RPA}
\end{equation}
compare with Eq.~\eqref{eqn: mean_field_chi_3}.

\subsubsection{Bubble}
The bare bubble can be computed directly as
\begin{gather}      
gB_0(\omega)=\vcenter{\hbox{\includegraphics[scale=0.8]{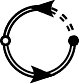}}}= \frac{2ig}{L^d}\sum_{\boldsymbol{k}}\int\frac{d\Omega}{2\pi}\mathcal{D}^R_{\boldsymbol{k}}(\omega+\Omega)\mathcal{D}^K_{\boldsymbol{k}}(\Omega).
\label{eqn: bubble_RPA}
\end{gather}
\subsubsection{Third-order response}
Collecting all the previously computed ingredients, the vertex correction is simply given by (note that the external legs have been amputated)
\begin{widetext}
\begin{equation}
    i\Gamma_{\rm RPA}(\omega_3,\omega_2,\omega_1) = \vcenter{\hbox{\includegraphics[scale=1]{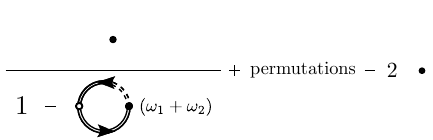}}}\quad,
\end{equation}
and we arrive directly at the third-order response
\begin{align}
    \chi^{(3)}_{\rm RPA}(\omega_1+\omega_2+\omega_3,\omega_1+\omega_2,\omega_1)= \underbrace{\left(2g-\sum_{i=1}^3\frac{g}{1- \, g B_0(\omega_1+\omega_2+\omega_3-\omega_i)}\right)}_{\displaystyle{i\Gamma_{\rm RPA}(\omega_3,\omega_2,\omega_1)}}\,\chi^{(1)}(\omega_1)\chi^{(1)}(\omega_2)\chi^{(1)}(\omega_3)\chi^{(1)}(\bar{\omega}),
    \label{eqn: RPA_chi_3}
\end{align}
\end{widetext}
with $\chi^{(1)}$ given by Dyson's equation, see Eq.~\eqref{eqn: Dyson}.
Note that for $B_0\ll1$, the RPA result asymptotically approaches the lowest order or ``mean field'' calculation, see Eq.~\eqref{eqn: mean_field_chi_3}. In the classical limit, $T\gg\epsilon_{\boldsymbol{k}}$, this result is the analogue of the sum of both ``mean field'' and ``squeezing'' contributions presented in~\cite{salvador2024principles}. As such, it serves as a generalization to the quantum limit and shows that the Gaussian state Ansatz for $\chi^{(3)}$ corresponds to a vertex correction within the RPA.

\subsection{2D maps within the RPA} \label{subsec: 2D_discussion_RPA}
\begin{figure*}
    \centering
    \includegraphics[width=0.75\linewidth]{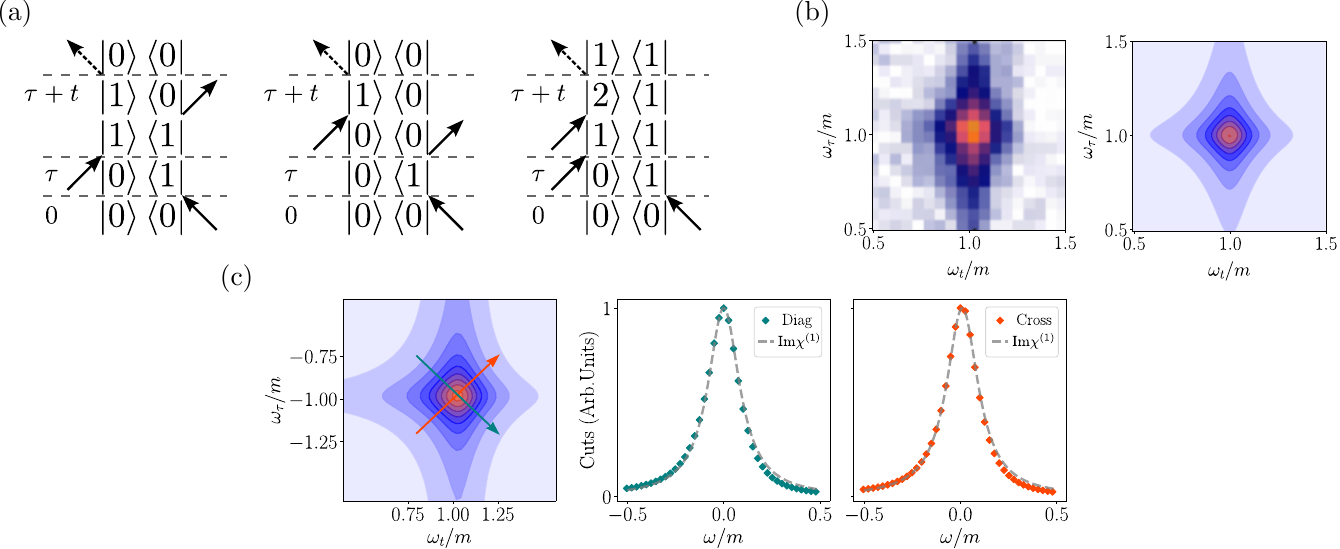}
    \caption{(a) Two-sided Feynman diagrams for bosonic excitations. The solid arrows indicate the interaction with the first and second pulse $E_A$ and $E_B$, while the dashed arrows represent the measurement. First two diagrams correspond to the ITLS-type paths, while the third is particular of bosonic excitations. Bosonic excitations can explore states beyond $\ket{0}$ and $\ket{1}$ resulting in asymmetric decay factors in $\tau$ and $t$, as discussed in Sec.~\ref{subsec: 2D_discussion_RPA}, and the possibility to end in a different state than the initial one after the protocol. (b) Nonrephasing peak experimentally obtained in Ref.~\cite{Liu_2023_echo} and theoretical nonrephasing peak from Eq.~\eqref{eqn: 2D_map_exact_eta}, with $\eta/m=0.1$, displaying the vertical elongation typical of bosons. (c) Theoretically obtained absolute-value echo peak and its cuts along the diagonal (Diag) and cross diagonal (Cross). Both cuts follow the the spectral function of the single bosonic excitation $\Im \chi^{(1)}$, in dashed gray, when the vertex corrections are small.}
    \label{fig: RPA_maps}
\end{figure*}
The tree level approximation of the nonlinear response, see Eq.~\eqref{eqn: mean_field_chi_3}, provides a reasonable approximation of the 2D map when the nonlinearity $g$ is small. Assuming a finite broadening $\eta$, the map is given by
\begin{widetext}
\footnotesize
\begin{equation}
    S(\omega_t,\omega_\tau) \sim \frac{2 (6 \eta -i (\omega_t+3 \omega_\tau ))}{(-i \eta +m-\omega_t) (-3 i \eta +m-\omega_t) (i \eta +m+\omega_t) (3 i \eta -3 m+\omega_t)(3 i \eta +m+\omega_t) (3 i \eta +3 m+\omega_t) (-i \eta +m-\omega_\tau ) (i \eta +m+\omega_\tau )}.
    \label{eqn: 2D_map_exact_eta}
\end{equation}
\normalsize
\end{widetext}
This expression provides valuable insights into key aspects of the 2D map for nonlinear collective excitations. First, the poles in Eq.~\eqref{eqn: 2D_map_exact_eta} exhibit a damping given by $\eta$ or $3\eta$, instead of the expected $\eta$ from the ITLS perspective. The appearance of these two different damping rates can be understood by tracing the different pathways contributing, for instance, to the echo response, see Fig.~\ref{fig: RPA_maps}(a) for a two-sided Feynman diagram interpretation. We identify the terms proportional to $e^{-\eta(\tau+t)}$ as the prototypical echo peaks of ITLS, corresponding to the first two paths in Fig.~\ref{fig: RPA_maps}(a). However, we also obtain an additional contribution proportional to $e^{-\eta(\tau+3t)}$, which corresponds to the last path in Fig.~\ref{fig: RPA_maps}(a). Note that this term contains a larger decay rate during the second time delay due to the fact that three excitations are coexisting during this time evolution. Since this extra path comes with a relative minus sign with respect to the previous two, adding up the three contributions results in an elongated peak along the $\tau$ axis (vertically), as can be seen in Fig.~\ref{fig: RPA_maps} (b), where we present the experimentally observed nonrephasing peak in~\cite{Liu_2023_echo} and the analytically obtained following Eq.~\eqref{eqn: 2D_map_exact_eta}. Such contribution is absent in the ITLS scenario, and the peaks are fully symmetric. 

Even more remarkable and of practical interest is the fact that if vertex corrections are moderate, the cuts along the diagonal and cross-diagonal of the echo peak coincide with each other and with the excitation linewidth, i.e., with the linear response broadening.
To illustrate this point, we plot in Fig.~\ref{fig: RPA_maps}(c) the absolute-value echo peak and its cuts as well as the imaginary part of the linear response function for the full RPA response, Eq.~\eqref{eqn: RPA_chi_3}, with moderate nonlinearity $g/m^2=0.1$. As can be seen, the overlap is remarkably accurate; note the absence of any fitting parameters in the comparison. We expect this feature to break down as soon as vertex corrections start to play an important role, since the nonlinear response directly depends on the vertex function; see Eq.~\eqref{eqn: general_diag}. One of the main contributions of the vertex is the inclusion of scattering corrections in the form of virtual quantum processes. Therefore, the 2D map contains relevant information regarding the interaction between the collective modes and it can reveal intriguing phenomena, such as bound state formation, the discussion of which will be presented elsewhere.

\section{Elastic Scattering}\label{sec: elastic}
This section provides a quantitative discussion of the computation of the third-order nonlinear response of collective excitations in the presence of disorder. The first subsection~\ref{subsection: diagram_echo_intuition}, is devoted to building an intuitive understanding of the necessary subclass of diagrams essential to capture the rephasing physics. In the second subsection~\ref{subsection: ladder_resumm}, we obtain the self-energy and full disorder vertex in the presence of disorder. Then, in the third subsection~\ref{subsec: chi_3_elastic}, we extend the diagrammatic calculation of $\chi^{(3)}$ to account for the disorder-induced corrections to the self-energy and vertex. 

We consider an elastic scattering term in the action given by
\begin{multline}
    S_{\rm dis} = -\int_{\mathcal{C}}dt\int d^d \boldsymbol{r}\, \frac{V(\boldsymbol{r})}{2}\varphi^2(\boldsymbol{r},t) \\= -\int_{-\infty}^{\infty} dt\int d^d \boldsymbol{r} \, 2V(\boldsymbol{r})\,\varphi^{cl}(\boldsymbol{r},t)\varphi^{q}(\boldsymbol{r},t),
    \label{eqn: S_dis}
\end{multline}
and assume a disorder distribution which is Gaussian and flat, i.e., $\overline{V} = 0$ and $\overline{V(\boldsymbol{r})V(\boldsymbol{r}')}=V_0^2\delta(\boldsymbol{r}-\boldsymbol{r}')$. Here, $V_0^2 = V^2 \xi^d$, where $V^2$ characterizes the strength of the disorder and $\xi$ its (short-range) correlation length. Integrating out the disorder results in a $\varphi$-dependent action, with an effective disorder induced vertex given by
\begin{multline}
    S_{\rm dis}^{\rm eff}[\varphi] =i\int dt \int dt'\int d^d\boldsymbol{r}\,2 V_0^2\\ \times \varphi^{cl}(\boldsymbol{r},t)\varphi^{q}(\boldsymbol{r},t)\varphi^{cl}(\boldsymbol{r},t')\varphi^{q}(\boldsymbol{r},t'). \\ 
    \label{eqn: S_dis_eff}
\end{multline}
The disorder vertex does not directly contribute to the non-linear response $\chi^{(3)}$ by its own due to its causality structure. Specifically, this implies that there is no leading term in $\chi^{(3)}$ solely proportional to $V_0^2$.

\subsection{Diagrammatic intuition of the echo peak} \label{subsection: diagram_echo_intuition}
\begin{figure}
    \centering
    \includegraphics[width=0.95\linewidth]{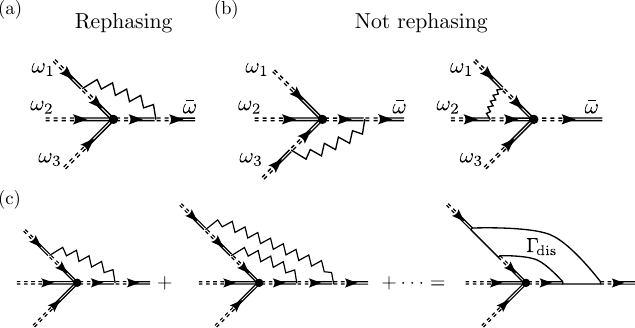}
    \caption{Intuition behind the diagrams for the echo peak. In (a) a disorder line correlates $\omega_1$, which corresponds to the first time oscillation, with $\bar{\omega}$, which corresponds to the second time delay oscillation. This results in the diagram presenting the rephasing structure. In (b) two examples of vertex corrections that do not correlate $\omega_1$ with $\bar{\omega}$, resulting in contributions that do not exhibit the rephasing structure. In (c) a nonperturbative ladder resummation for non-crossed disorder lines connecting modes with frequencies $\omega_1$ and $\bar{\omega}$.}
    \label{fig: echo_intuition}
\end{figure}
We expand here the intuitive discussion of Sec.~\ref{Sec: intuition} by adapting it to the language of diagrammatics, and justify the subclass of diagrams necessary to capture the echo signature. As previously argued, the origin of the almond-like structure of the echo peak relies on the system evolving with opposite frequencies during time delays $\tau$ and $t$. Following the notation set by Eq.~\eqref{eqn: S_t_tau_B}, the first (second) time delay $\tau$ ($t$) is dominated by the dynamics of $\omega_1$ ($\bar{\omega}$). This can intuitively be understood since during the first time delay, only the first pulse ($\omega_1$) has interacted with the system, while for the second time delay, the effect of both pulses ($\bar{\omega}$) is present. Combining this with the diagrammatic representation in Eq.~\eqref{eqn: general_diag}, we deduce that the diagrams that have a rephasing nature are the ones correlating propagators with frequency $\omega_1$ and $\bar{\omega}$, see Fig.~\ref{fig: echo_intuition}(a). However, there are other diagrams which lack this rephasing structure, specifically those that correlate propagators other than the ones with frequencies $\omega_1$ and $\bar{\omega}$, see Fig.~\ref{fig: echo_intuition}(b) for two examples. 

As we discuss in Sec.~\ref{subsec: chi_3_elastic}, in order to capture the effects of disorder, it is not enough to consider the lowest order disorder corrections. This is due to the intrinsically nonperturbative character of disorder effects~\cite{sadovskii2006diagrammatics}. In this work, we consider the ladder resummation of the previously discussed class of diagrams, i.e., the infinite series of noncrossing disorder lines, see Fig.~\ref{fig: echo_intuition}(c). 
By considering the infinite set of disorder lines connecting the first and last excitation, we enhance the correlations between them which results in a dramatic increase of the rephasing. However, the infinite resummation of non-crossed disorder lines connecting other modes does not lead to sharp signatures in the echo peak. In the next subsection, we proceed to perform the ladder resummation of the disorder effects without considering nonlinearity. 

A second family of diagrams involving impurity scattering crossings can be considered beyond the ladder approximation. However, similar arguments to the fermionic counterpart can be applied to safely disregard them in the limit of $E \tau_e \gg 1$~\cite{sadovskii2006diagrammatics,akkermans2007mesoscopic}, where $E$ is the typical excitation energy and $\tau_e$ the elastic collision time, which can be associated to the decay time of the single-boson propagator. The typical energy scale can be associated to the boson mass $E\sim m$, and the typical collision time $\tau_e^{-1}~\sim V_0^2/v^dm$ can be obtained from the Born approximation to the impurity scattering, see Eq.~\eqref{eqn: self_energy_born}. The validity of the ladder approximation thus relies on $ (m v^{d/2}/V_0)^2\gg 1$.

\subsection{Ladder resummation of disorder}\label{subsection: ladder_resumm}
\begin{figure}
    \centering
    \includegraphics[width=0.95\linewidth]{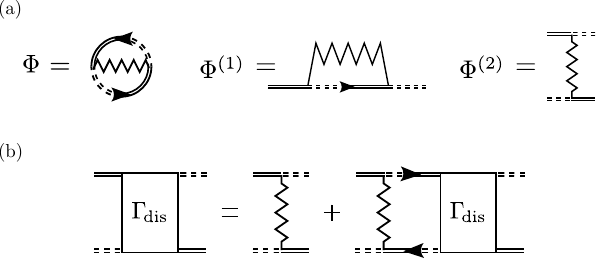}
    \caption{(a) Luttinger-Ward functional approximation to lowest order in disorder. (b) Bethe-Salpeter equation for the disorder vertex.}
    \label{fig: LW_disorder}
\end{figure}
The Luttinger-Ward functional for the noninteracting disordered system is depicted in Fig.~\ref{fig: LW_disorder}. By performing a derivative of the functional with respect to $\mathcal{D}^R(\omega)$ we obtain the self-consistent self-energy within the Born (noncrossing) approximation ($v=1$ henceforth),
\begin{align}
    \Sigma_V(\omega) &= \vcenter{\hbox{\includegraphics[scale=0.65]{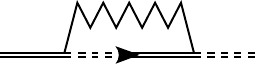}}} = \frac{4 V_0^2}{L^d}\sum_{\boldsymbol{k}}\mathcal{D}^R_{\boldsymbol{k}}(\omega). 
    \label{eqn: self_energy_born}
\end{align} 
In principle, this equation has to be solved self-consistently for the real and the imaginary part. However, the real part corresponds to an unimportant shift and therefore restrict ourselves to the imaginary part.
It is worth noting the following relation of the self-consistent equation:
\begin{equation}
    \text{Im}\Sigma_V(\omega) = \frac{4V_0^2}{L^d}\sum_{\boldsymbol{k}}\abs{\mathcal{D}_{\boldsymbol{k}}^R(\omega)}^2\text{Im}\Sigma_{\boldsymbol{k}}(\omega).
    \label{eqn: Born_general}
\end{equation}
Assuming a momentum independent total self-energy $\Sigma(\omega)$,
we can further simplify the self-consistent equation to
\begin{equation}
    \frac{\text{Im}\Sigma_V(\omega)}{\text{Im}\Sigma(\omega)} =\frac{4V_0^2}{L^d}\sum_{\boldsymbol{k}}\abs{\mathcal{D}_{\boldsymbol{k}}^R(\omega)}^2.
    \label{eqn: self_consistent_Born}
\end{equation}

%

Having obtained the disorder contribution to the self-energy, we proceed to compute vertex correction in the $cl-q-cl-q$ channel. The bare disorder vertex $\Gamma^{(0)}_{\rm dis}$ can be obtained within the Baym-Kadanoff formalism by taking a functional derivative of the self-energy,
\begin{equation}    
    \Gamma^{(0)}_{\rm dis}(\boldsymbol{k};\omega_a,\omega_b) = \frac{\delta \Sigma_V(\omega_a)}{\delta \mathcal{D}^R_{\boldsymbol{k}}(\omega_b)}=\vcenter{\hbox{\includegraphics[scale=0.65]{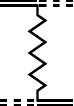}}}.
    \label{eqn: bare_disorder_vertex}
\end{equation}
Finally, to obtain the full disorder vertex $\Gamma_{\rm dis}$ we solve the Bethe-Salpeter equation which is depicted in Fig.~\ref{fig: LW_disorder}(b),
\begin{equation}  
    \Gamma_{\rm dis}(\boldsymbol{k};\omega_a,\omega_b) = \frac{2V_0^2}{1-V_0^2 \lambda(\boldsymbol{k};\omega_a,\omega_b)},
\end{equation}
where we have introduced the disorder bubble:
\begin{equation}
\lambda(\boldsymbol{k};\omega_a,\omega_b)=\frac{4}{L^d}\sum_{\boldsymbol{q}} \mathcal{D}_{\boldsymbol{k}+\boldsymbol{q}}^R(\omega_a)\mathcal{D}_{\boldsymbol{q}}^R(\omega_b).
\label{eqn: disorder_bubble}
\end{equation}
For $\boldsymbol{k}=0$ and under the assumption of a momentum independent self-energy, we can obtain an important exact relation between the disorder bubble and the self-energy. First, we rewrite $\lambda(\boldsymbol{0};\omega_a,\omega_b)$ by means of a simple fraction decomposition as
\begin{equation}
    \lambda(\boldsymbol{0};\omega_a,\omega_b) = \frac{\frac{4}{L^d}\sum_{\boldsymbol{q}}\mathcal{D}^R_{\boldsymbol{q}}(\omega_a)-\frac{4}{L^d}\sum_{\boldsymbol{q}}\mathcal{D}^R_{\boldsymbol{q}}(\omega_b)}{\Sigma(\omega_a)-\Sigma(\omega_b)-2\omega_a^2+2\omega_b^2},
\end{equation}
and assuming there exists a self-consistent solution to  Eq.~\eqref{eqn: self_energy_born} we arrive at
\begin{equation}
    \lambda(\boldsymbol{0};\omega_a,\omega
    _b) = \frac{1}{V_0^2}\frac{\Sigma_V(\omega_a)-\Sigma_V(\omega_b)}{\Sigma(\omega_a)-\Sigma(\omega_b)-2\omega_a^2+2\omega_b^2}.
\end{equation}
In particular, and of special interest for the upcoming section, for $\omega_a=-\omega_b=\omega$ it simplifies to
\begin{equation}
    \lambda(\boldsymbol{0};\omega,-\omega) =\frac{1}{V_0^2} \frac{\text{Im}\Sigma_V(\omega)}{\text{Im}\Sigma(\omega)}.
    \label{eqn: vertex_self_energy}
\end{equation}
Note that the limit $\omega_a=\omega_b=\omega$ is also well-defined, and by means of Eq.~\eqref{eqn: self_energy_born} can be found to be determined by
\begin{equation}
    \lambda(\boldsymbol{0};\omega,\omega) = \frac{4}{L^d}\sum_{\boldsymbol{q}}\left[\mathcal{D}^R_{\boldsymbol{q}}(\omega)\right]^2,
    \label{eqn: vertex_other_diag}
\end{equation}
in agreement with Eq.~\eqref{eqn: disorder_bubble}.

%

Having obtained the RPA resummation without disorder and the full disorder vertex without interaction, we proceed to compute the effects of disorder on the nonlinearity.

\subsection{\texorpdfstring{$\chi^{(3)}$}{x(3)} in the presence of disorder} \label{subsec: chi_3_elastic}
\begin{figure}
    \centering
    \includegraphics[width=0.8\linewidth]{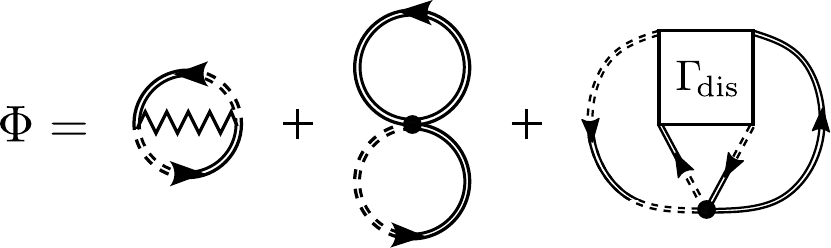}
    \caption{Luttinger-Ward functional employed for the computation of $\chi^{(3)}$ in the presence of elastic scattering. In addition to the RPA contribution and the disorder contribution, the last term corresponds to an interaction-disorder term.}
    \label{fig: LW_full}
\end{figure}
To capture the echo physics it is essential to include disorder corrections to the interaction vertex, as discussed in Sec.~\ref{Sec: intuition}. This is achieved by introducing an additional disorder-interaction contribution to the Luttinger-Ward functional within the non-crossing approximation, illustrated in Fig.~\ref{fig: LW_full}. Specifically, we incorporate the infinite series of noncrossing disorder diagrams into the functional by using $\Gamma_{\rm dis}$ instead of $\Gamma^{(0)}_{\rm dis}$. Notably, the $\chi^{(3)}$ obtained when considering only $\Gamma^{(0)}_{\rm dis}$ does not exhibit the almond-like signature in the echo peak, although it provides an instructive hint toward the correct physical mechanism. For completeness, the details of this calculation, along with the resulting 2D map signatures, are presented in Appendix~\ref{Sec: tree_level_disorder}. The necessity to consider $\Gamma_{\rm dis}$ underscores the non-perturbative nature of the echo peak signature in relation to disorder and highlights the potential of 2D spectroscopy as a tool for probing more complex disorder effects. Following the same structure as in Sec.~\ref{sec: RPA} we proceed to evaluate the necessary diagrams to obtain $\chi^{(3)}$ in the presence of disorder. 

\subsubsection{Self-energy}
The interaction and disorder-only contributions to the self-energy arising from the Luttinger-Ward functional are given by Eqs.~\eqref{eqn: self_energy_RPA} and~\eqref{eqn: self_energy_born} respectively.
There are two additional contributions coming from the disorder-interaction part:
\begin{multline} 
    \Sigma_{g-V}(\omega,\boldsymbol{k})=\vcenter{\hbox{\includegraphics[scale=0.75]{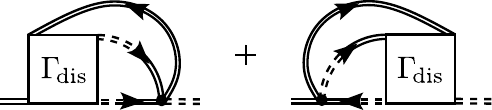}}}\\=2\times\frac{4ig}{L^{2d}}\sum_{\boldsymbol{k},\boldsymbol{q}}\int\frac{d\Omega}{2\pi}\,\mathcal{D}^R_{\boldsymbol{p}}(\omega)\mathcal{D}^R_{\boldsymbol{p}+\boldsymbol{q}-\boldsymbol{k}}(\Omega)\mathcal{D}^K_{\boldsymbol{q}}(\Omega)\\ \times \Gamma_{\rm dis}(\boldsymbol{q}-\boldsymbol{k};\omega,\Omega).
    \label{eq: selfenergy_elastic}
\end{multline}
Furthermore, we include a bath-like term in the self-energy of the form $\Sigma_{bath}(\omega)=-i2\omega\gamma$ to capture the effects of inelastic scattering.
%
%
\subsubsection{Vertices}
The bare vertices contributing to the nonlinear response are given by taking a derivative with respect to the Keldysh Green's function and setting the external momenta to $\boldsymbol{0}$. The RPA contribution is given in Eq.~\eqref{eqn: bare_vertex_RPA} and the interaction-disorder contribution to the Luttinger-Ward results in the following vertices:
\begin{align}
    \vcenter{\hbox{\includegraphics[scale=0.6]{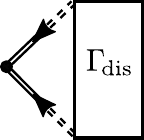}}} = &ig\left(\frac{4}{L^d}\sum_{\boldsymbol{k}} \mathcal{D}_{\boldsymbol{k}}^R(\omega_a)\mathcal{D}_{-\boldsymbol{k}}^R(\omega_b)\right)\Gamma_{\rm dis}(\boldsymbol{0};\omega_a,\omega_b)  \notag \\
    =&ig\frac{2V_0^2\lambda(\boldsymbol{0};\omega_a,\omega_b)}{1-V_0^2\lambda(\boldsymbol{0};\omega_a,\omega_b)} \equiv\mathcal{T}_{g-V}(\omega_a,\omega_b), \notag \\
    \vcenter{\hbox{\includegraphics[scale=0.65]{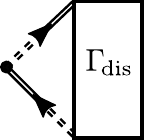}}}=&ig\left(\frac{4}{L^d}\sum_{\boldsymbol{k}} \mathcal{D}_{\boldsymbol{k}}^R(\omega_a)\mathcal{D}_{\boldsymbol{k}}^R(\omega_b)\right)\Gamma_{\rm dis}(\boldsymbol{0};\omega_a,\omega_b) \notag \\
    = &ig\frac{2V_0^2\lambda(\boldsymbol{0};\omega_a,\omega_b)}{1-V_0^2\lambda(\boldsymbol{0};\omega_a,\omega_b)} \equiv\mathcal{T}_{g-V}(\omega_a,\omega_b).
    \label{eqn: elastic_vertex_correction}
\end{align}
Note that we have employed the inversion symmetry of our system, $\epsilon_{\boldsymbol{k}}=\epsilon_{-\boldsymbol{k}}$. In principle, inelastic scattering processes could contribute to the vertex of the collective mode. However, vertex corrections can be safely disregarded in two different scenarios: when the collective mode is coupled to another massive collective mode with a significantly larger mass, and when it is coupled to a continuum of quasiparticles with a density of states sharply peaked at energies higher than the mass. In Appendix~\ref{sec: inelastic}, we provide a microscopic model for the former and demonstrate the irrelevance of vertex corrections in this situation.

\subsubsection{Bubbles}
The bare bubble arising from the RPA contribution is given by Eq.~\eqref{eqn: bubble_RPA}. The additional interaction-disorder term gives rise to a second bubble given by:
\begin{widetext}
\begin{align} 
    g B_{\rm dis}(\omega)=\vcenter{\hbox{\includegraphics[scale=0.7]{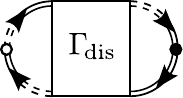}}}
    &= 2ig\int \frac{d\Omega}{2\pi}\left(\frac{1}{L^d}\sum_{\boldsymbol{q}}\mathcal{D}_{\boldsymbol{q}}^{R}(\omega+\Omega)\mathcal{D}_{\boldsymbol{q}}^{K}(\Omega)\right)\left(\frac{4}{L^d}\sum_{\boldsymbol{q}}\mathcal{D}_{\boldsymbol{q}}^R(\omega+\Omega)\mathcal{D}_{\boldsymbol{q}}^R(\Omega)\right)\Gamma_{\rm dis}(\boldsymbol{0};\omega+\Omega,\omega)
    \notag \\ &= 2ig\int \frac{d\Omega}{2\pi}\left(\frac{1}{L^d}\sum_{\boldsymbol{q}}\mathcal{D}_{\boldsymbol{q}}^{R}(\omega+\Omega)\mathcal{D}_{\boldsymbol{q}}^{K}(\Omega)\right)\frac{2V_0^2\lambda(\boldsymbol{0};\omega+\Omega,\Omega)}{1-V_0^2\lambda(\boldsymbol{0};\omega+\Omega,\Omega)}.
    \label{eq: bubble_elastic}
\end{align}
This concludes the calculation of all the necessary ingredients to compute $\chi^{(3)}$ in the presence of disorder within our non-crossing approximation.

\subsubsection{Third-order response}
Collecting all the pieces together, the vertex $i\Gamma$ is diagrammatically given by

\begin{equation}
    i\Gamma(\omega_3,\omega_2,\omega_1) = \vcenter{\hbox{\includegraphics[scale=0.575]{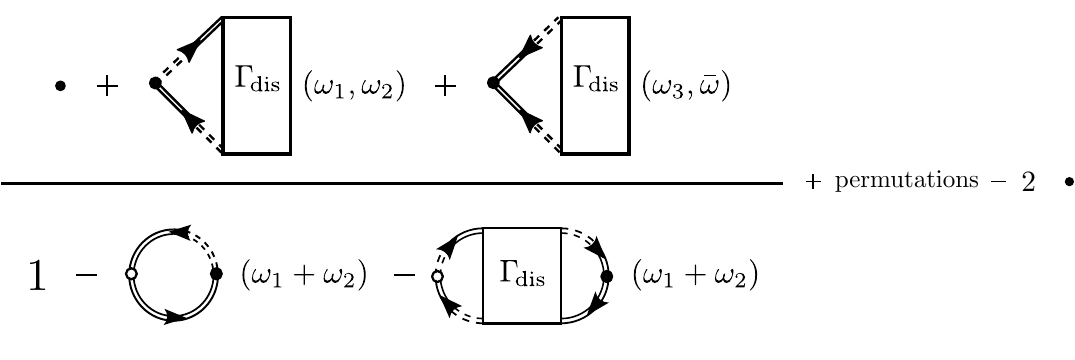}}}\quad,
    \label{eq: vertex_elastic}
\end{equation}
where permutations refer to unique pairs without accounting for the frequency ordering. In total there exist three of such permutations.
\end{widetext}

\section{Weak nonlinear regime}\label{sec: classical_limit}
\begin{figure}
    \centering
    \includegraphics[width=1\linewidth]{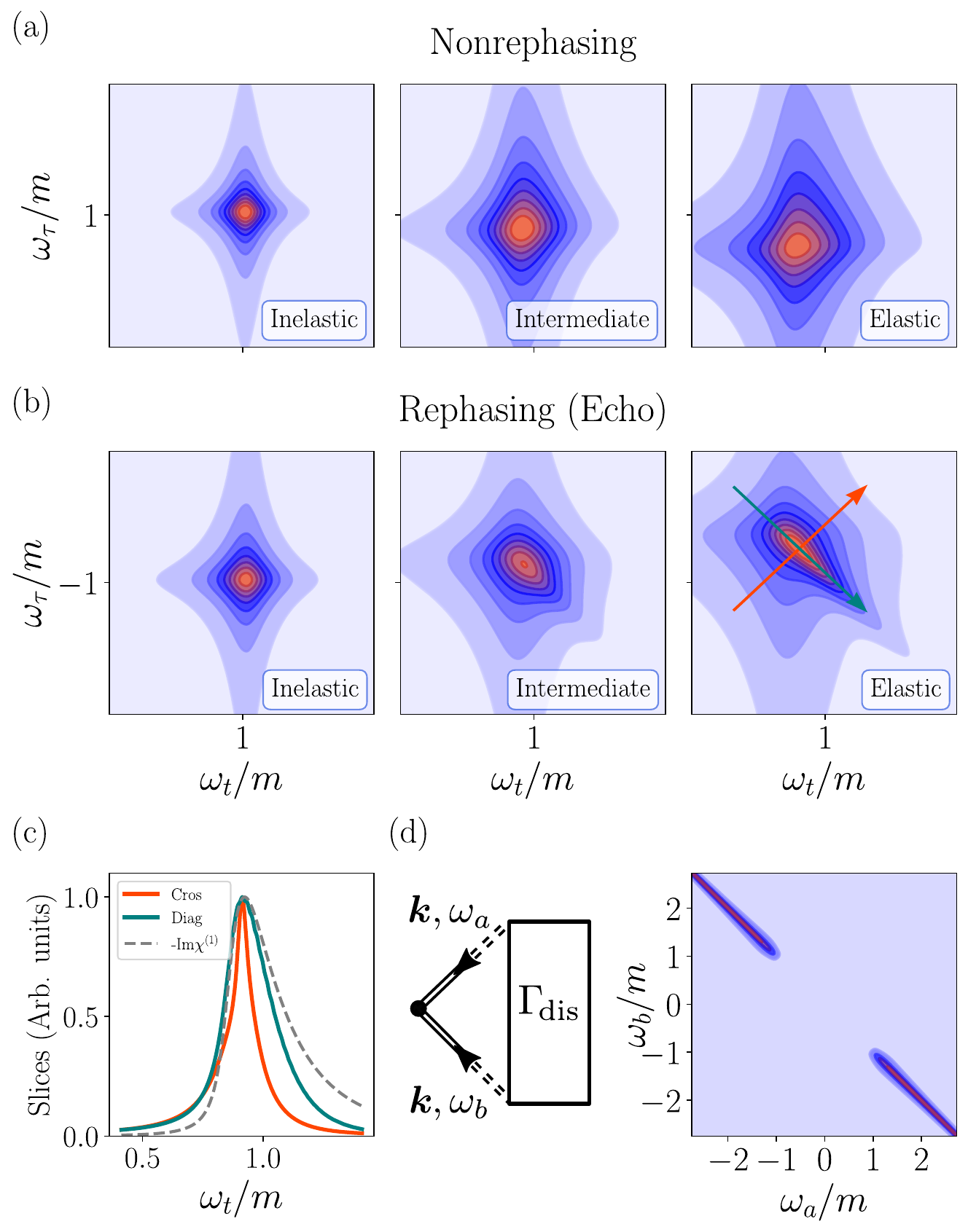}
    \caption{Weak nonlinear regime of the 2D map in the presence of disorder. (a)-(b) Evolution of the Nonrephasing and Rephasing (Echo) peaks for the small $g$ limit in the inelastic, intermediate, and elastic regimes, with values $V_0^2/m^2=\{0,1.5,3\}$ and $\gamma/m=\{0.1,0.075,0.05\}$ respectively. The Nonrephasing peak (a) becomes broad in the presence of disorder while the Echo peak (b) displays the characteristic almond-like signature signaling the presence of elastic scattering. Note that the slight drift of the peak to lower frequencies is a consequence of the dampening induced by the disorder. (c) Slices along the diagonal (teal) and cross-diagonal (orange) of the echo peak. The dashed gray line corresponds to the spectral function $-\Im \chi^{(1)}$. (d) Diagram and color-plot of the absolute value of the tree-level vertex correction in the presence of disorder. The vertex correction is most prominent along the $\omega_a=-\omega_b$ line heralding the rephasing nature of the process. All the plots are obtained in the $d=2$ scenario.}
    \label{fig: classical}
\end{figure}
With all the necessary ingredients for computing the nonlinear response of a collective excitation in a disordered many-body system, we now proceed to study the effects of disorder on the 2D map in the weak nonlinear regime ($g/m^2\ll1$). In this limit, self-energy and bubble diagrams that couple disorder and nonlinearity can be disregarded, see Eqs. \eqref{eq: selfenergy_elastic} and \eqref{eq: bubble_elastic}. These processes lead to virtual frequency transfers and correspond to quantum fluctuations of the vacuum coupled to the disorder potential. Their effects on the 2D map will be discussed in Sec. \ref{sec: quantum_limit}. 

In agreement with the intuition put forward in Sec.~\ref{Sec: intuition}, we find from a quantitative perspective that the central ingredient to capturing rephasing effects in the echo peak lies in the vertex corrections induced by the static disorder $\mathcal{T}_{g-V}(\omega_a,\omega_b)$. In Fig.~\ref{fig: classical}(d) we present a color plot of the absolute-value disorder-corrected vertex. As can be directly seen, the contribution is most notable in the vicinity of $\omega_a=-\omega_b$ and for $\abs{\omega_a}$, $\abs{\omega_b}>m$. This asymmetry between the two diagonals, favoring opposite sign frequencies, heralds the echo physics contained in the vertex correction and, in turn, is responsible for the appearance of the characteristic almond-shaped peak in the echo nonlinearity. Setting $\omega_a=-\omega_b = \omega$ in Eq.~\eqref{eqn: elastic_vertex_correction}, the importance of relation Eq.~\eqref{eqn: vertex_self_energy} becomes evident. For the current discussion we have $\text{Im}\Sigma(\omega) = \text{Im}\Sigma_V(\omega)+\text{Im}\Sigma_{bath}(\omega)$, with $\Sigma_{bath}(\omega)=-i2\omega\gamma$, and we find for $\mathcal{T}_{g-V}$, defined in Eq.~\eqref{eqn: elastic_vertex_correction},
\begin{equation}
    \mathcal{T}_{g-V}(\omega,-\omega) = 2ig\frac{\text{Im}\Sigma_V(\omega)}{\text{Im}\Sigma_{bath}(\omega)} = ig\frac{-\text{Im}\Sigma_V(\omega)}{\omega\gamma}.
    \label{eqn: vertex_self_energy_classical}
\end{equation}
This is the central analytical result of this section, which highlights, in perfect agreement with the ITLS intuition, all the expected trends as a function of the ratio $V_0^2/m\gamma$. Note that Eq.~\eqref{eqn: vertex_self_energy_classical} is, given our form of the self-energy, exact within perturbation theory in $g$ in all spatial dimensions. We discuss in the following the inelastic ($V_0^2/m\gamma\ll1$) and elastic ($V_0^2/m\gamma\gg1$) limits.


In the inelastic scattering dominated regime, we can treat the imaginary part of the Born self-energy perturbatively. Since the self-energy is bounded, $\mathcal{T}_{g-V}$ scales as $V_0^2/m\gamma$ and thus vanishes as $V_0^2/m\gamma\rightarrow0$. Consequently, the disorder-induced vertex corrections vanish and the echo peak remains symmetric, as discussed in Section~\ref{sec: RPA}. Note that even in the limit of $\omega\rightarrow0$, since $-\text{Im}\Sigma_V(\omega)\sim V_0^2\omega\gamma/m^{4-d}$, $\mathcal{T}_{g-V}$ is finite and thus vanishes with $V_0^2\rightarrow0$, as expected. 

In the elastic limit, $\mathcal{T}_{g-V}(\omega,-\omega)$ diverges as $V_0^2/\omega\gamma$ and the 2D map displays a line of singularities at the echo diagonal, i.e. perfect rephasing. Intuitively, such a divergence can only occur for frequencies larger than the mass, where the system can oscillate, and can be confirmed by looking at the support of the disorder corrected vertex, see Fig.~\ref{fig: classical}(d). This divergent behavior bears an interesting connection with the physics of electrons and light in disordered media. Specifically, we are encountering the analogous of the diffuson pole at zero momentum for the collective excitation~\cite{sadovskii2006diagrammatics,akkermans2007mesoscopic,bruus2004many}.

Having discussed both limits, we plot in Fig.~\ref{fig: classical}(a) and (b) the evolution of the nonrephasing and rephasing (echo) peaks in the inelastic, intermediate, and elastic regimes. While the nonrephasing peak only becomes broader following the excitation linewidth, the echo peak exhibits the characteristic almond shape, qualitatively analogous to the ITLS signature and demonstrating the ability of the echo peak to distinguish between elastic and inelastic processes in a classical many-body system. An important difference with respect to the ITLS scenario lies in the asymmetry of the almond-like signature across the diagonal. In the ITLS case, the Larmor frequencies are typically assumed to follow a Gaussian distribution, leading to a symmetric echo peak with respect to reflections along the cross-diagonal, c.f. Fig.~\ref{fig: intro_echoes}. In contrast, for collective excitations driven at zero momentum, there are no finite momentum states below the mass available to be scattered into via the static disorder. Consequently, the explored virtual states always have higher energies, resulting in an almond shape which is elongated exclusively towards higher energies, c.f. Fig.~\ref{fig: classical}(b). This becomes especially evident when considering the slices of the echo peak across both diagonals, see Fig.~\ref{fig: classical}(c). We note that this asymmetry is compatible with the echo peak experimentally measured in~\cite{Liu_2023_echo}. Furthermore, we also plot in Fig.~\ref{fig: classical}(c) the spectral function in dashed gray and note that it deviates from the diagonal slices of the 2D map. This is in stark contrast to the discussion in~\ref{subsec: 2D_discussion_RPA}, where in the absence of strong vertex corrections both diagonal and cross diagonal cuts are equivalent and coincide with the spectral function.


\section{Strong nonlinear regime} \label{sec: quantum_limit}
Having established the connection between the standard ITLS intuition and the weak nonlinear regime of our theory, we move on to the study of the interaction induced quantum fluctuations, which are naturally present in our Luttinger-Ward functional. These include self-energy corrections beyond the Born approximation and bubble resummations arising from solving the Bethe-Salpeter equation. Numerical evaluation of the bubble corrections turn out to be parametrically small and only relevant for $g/m^2\sim V_0^2/m^2\sim 10$, and therefore can be completely disregarded at this level of approximation. On the other hand, we find that corrections arising from the self-energy are of utmost importance for the echo physics in the many-body scenario.

In the small $g$ limit we have discarded the interaction induced quantum fluctuations arising from the self-energy in Eq.~\eqref{eq: selfenergy_elastic}.  We want to understand how this corrections modify the discussion of the previous section. Note that, in our case, this additional self-energy term is a mixed contribution between the disorder and the nonlinearity $\Sigma_{g-V}~\sim gV_0^2$, and it is a function of momentum. Focusing again on $\omega_a=-\omega_b = \omega$ and combining Eqs.~\eqref{eqn: Born_general} and~\eqref{eqn: elastic_vertex_correction}, we generalize Eq.~\eqref{eqn: vertex_self_energy_classical} to arrive at
\begin{equation}
    \mathcal{T}_{g-V}(\omega,-\omega) = 2ig\frac{-\text{Im}\Sigma_V(\omega) + \mathcal{Q}(\omega)}{-\text{Im}\Sigma_{bath}(\omega) - \mathcal{Q}(\omega) },
    \label{eqn: vertex_self_energy_quantum}
\end{equation}
where $\mathcal{Q}(\omega) = 4V_0^2\sum_k\abs{\mathcal{D}_k(\omega)}^2\text{Im}\Sigma_{g-V}(k,\omega)$ represents the interaction induced quantum fluctuations arising from a momentum dependent self-energy. As previously mentioned, $\mathcal{Q}$ has only one contribution due to the choice of our Luttinger Ward functional, but Eq.~\eqref{eqn: vertex_self_energy_quantum} is general, with $\mathcal{Q}(\omega)$ containing all possible momentum dependent self-energy corrections. 

\begin{figure}
    \centering
    \includegraphics[width=\linewidth]{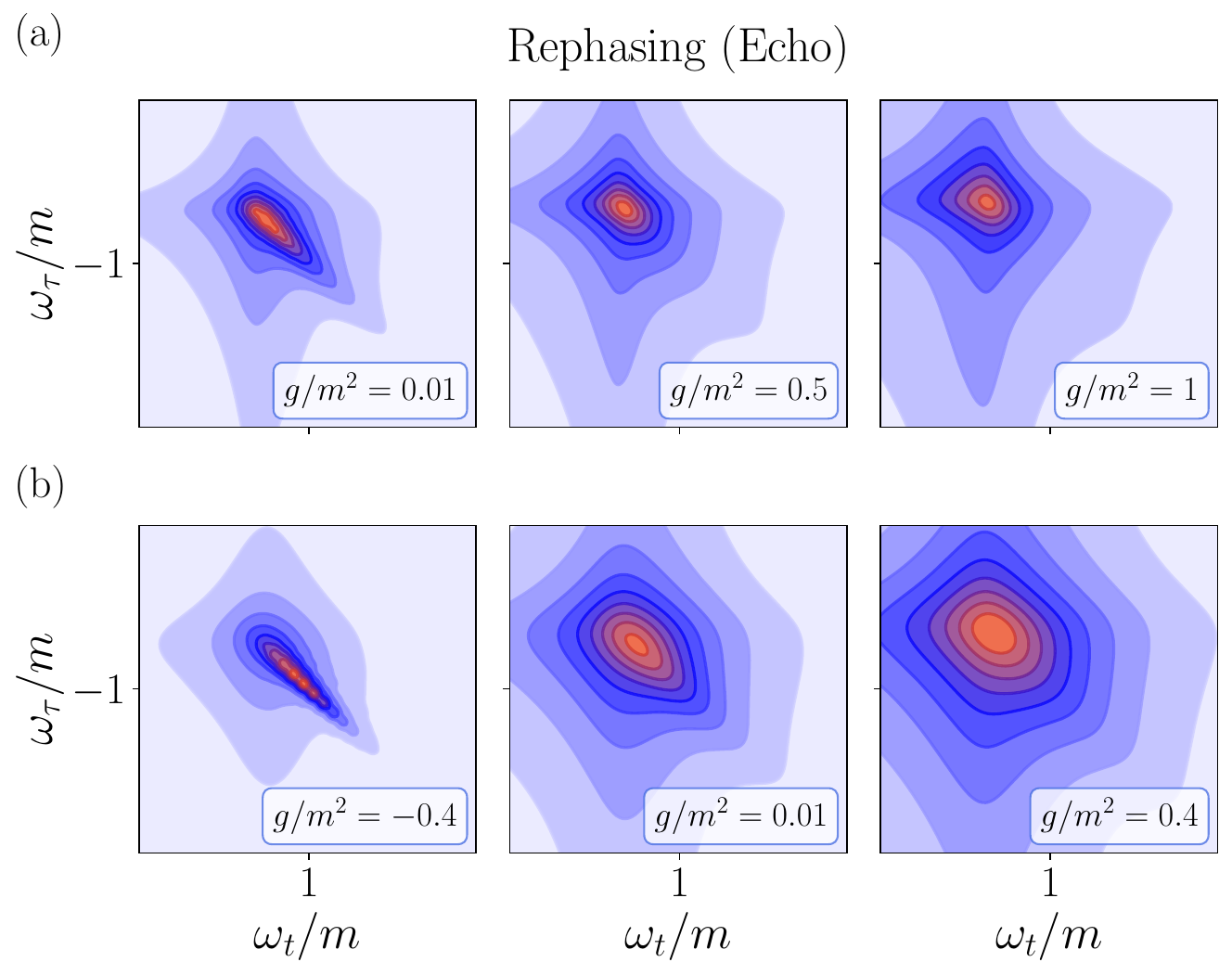}
    \caption{Echo peak in the presence of interaction induced quantum fluctuations. (a) Suppression of the echo peak due to the increase of $g$ for $V_0^2/m^2=3$ and $\gamma/m=0.01$. Quantum fluctuations arising from interactions act as an additional source of dephasing which cannot be rephased and destroy the almond-shape of the echo peak. (b) Quantum fluctuation enhancement (left panel) and suppression (right panel) of the echo peak as a function of the sign of $g$, for $V_0^2/m^2=3$ and $\gamma/m=0.075$. All the plots are obtained in the $d=2$ scenario.}
    \label{fig: quantum_corrections}
\end{figure}

The first consequence of Eq.~\eqref{eqn: vertex_self_energy_quantum} is the fact that quantum fluctuations due to the nonlinear nature of the excitations give rise to an intrinsic dephasing which cannot be rephased via echo type experiments. Even in the absence of an inelastic bath ($\Sigma_{bath}=0$) there is no perfect rephasing since there is no true divergence in Eq.~\eqref{eqn: vertex_self_energy_quantum} when $\mathcal{Q}(\omega)\neq0$. This can be intuitively understood, since the self-energy in Eq.~\eqref{eq: selfenergy_elastic} can be thought of as an effective inelastic scattering for the excitation, i.e. with a simultaneous transfer of energy and momenta. In the ITLS analogy, this energy fluctuation corresponds to having a fluctuating $B$-field, with a resulting linewidth broadening which is not possible to remove via echo-type processes. We present in Fig.~\ref{fig: quantum_corrections}(a) the evolution of a sharp echo peak as interactions increase, considering $V_0^2/m^2=3$ and $\gamma/m=0.01$ ($\gamma$ is kept finite due to resolution limitations). As can be seen, increasing $g$ has a similar effect as increasing $\gamma$ such that it effectively broadens the peak and softens the almond-like signature, c.f. Fig.~\ref{fig: classical}(b).

Secondly, in the presence of a bath, interaction induced quantum fluctuations can enhance or suppress the echo response, depending on the sign of $\mathcal{Q}(\omega)$. For example, in our case, $\mathcal{Q(\omega)}$ depends linearly on $g$, such that its sign determines whether $\mathcal{T}_{g-V}(\omega,-\omega)$ sharpens or broadens. In particular, for $g>0$, the rephasing is suppressed while for $g<0$ it is enhanced. Intuitively, two excitations traveling through a disordered media experience a larger correlated scattering from the impurities if they are spatially closer together. The evolution of the echo peak for different values of $g$ is presented in Fig.~\ref{fig: quantum_corrections} for $V_0^2/m^2=3$ and $\gamma/m = 0.075$. In the center panel we consider a perturbative $g$, such that $\mathcal{Q}(\omega)$ is negligible and we recover the weak nonlinear regime. On the right (left) panel we consider the echo peak with a finite positive (negative) seizable value of $g$ displaying a quantum-fluctuation-induced suppression (enhancement) of the echo peak.

\section{Conclusions and Outlook}\label{sec: conclusions}

In this paper, we have introduced a new framework for understanding disorder effects in the nonlinear response of collective many-body bosonic excitations. Specifically, we have shown that the standard treatment used for interpreting two-dimensional spectroscopy, originally developed for isolated two-level systems, breaks down in the presence of many-body correlations. 
We have developed a quantitative field-theoretical approach to calculate the third-order nonlinear response in the presence of disorder within the Keldysh formalism, and employed it to compute the spectra of two-dimensional spectroscopy of collective excitations. 
Through a non-perturbative treatment of static disorder, we have demonstrated that the echo peak exhibits an almond-like signature which heralds the presence of elastic scattering.
In the limit of small nonlinearities, we have recovered the standard ITLS phenomenology of the echo peak shape that arises from the interplay between elastic and inelastic processes, and quantified it in terms of their respective self-energy contributions.
Furthermore, we have demonstrated that as interactions are increased, strong quantum fluctuations can significantly alter the shape of the echo peak and, in particular, act as a source of dephasing which prevents perfect rephasing to be attainable in many-body systems.

Of particular interest, and serving both as the motivation and inspiration for this work, is the study of the effects of spatial inhomogeneity in the superfluid density on the Josephson plasmon excitations of high temperature superconducting cuprates. Specifically, the recent experimental realization of two-dimensional terahertz spectroscopy for the Josephson plasmon of La$_{2-x}$Sr$_x$CuO$_4$ near optimal doping has unearthed intriguing features of the echo peak as a function of temperature~\cite{Liu_2023_echo}. These results offer an ideal platform for applying our theoretical developments to extract valuable insights into the nature and effects of disorder in cuprates~\cite{Salvador_2025_cuprate_echo_thory}.

We envision that our work on two-dimensional spectroscopy of collective bosonic excitations can serve as a foundation for investigating several intriguing open questions. A natural extension of this work would be to explore the impact of real space localization of collective excitations on the echo peak. In this limit, the standard treatment used in isolated two-level systems should become accurate, allowing for a perfect rephasing and potentially serving as a probe of localization.
Of particular interest could be the study of the nonlinear response of disordered Bose-Einstein condensates, where Anderson localization has been observed in the past~\cite{Billy2008,Roati2008}. Nonlinear spectroscopy could provide a new framework for understanding the impact of nonlinearities on disordered systems, potentially leading to new insights into many-body localization~\cite{Mahmood2021}.
Similarly to how multidimensional-coherent spectroscopy probes biexcitons and trions~\cite{stone2009two,hao2017trion,hao2017neutral}, two-dimensional terahertz spectroscopy could be envisioned as a tool to investigate bound states of bosonic collective excitations, for instance biphonons or biplasmons. Extending our approach to fifth-order nonlinearities would capture the rephasing of two-quantum coherences and open avenues for studying the effects of disorder on bound states.~\cite{Fulmer_fifth_order_bound}.
Furthermore, inspired by the capabilities of two-dimensional spectroscopy, a promising direction would be to study the coupling and hybridization between different fields. 
For instance, extending the current analysis to models with several coupled low-energy excitation could elucidate unique features of interacting fields.

Recent advancements have shown that dynamic manipulation of quantum materials using light can induce non-trivial correlated phases~\cite{Colloquium2021UltrafastControl}, such as light-induced ferroelectricity~\cite{Li2019LightInducedFerro,Nova2019LightInducedFerro} or superconductivity~\cite{Fausti2011LightInducedSupercond,Mitrano2016LightInducedSuper,Fava2024YBCOExpulsion} among many others. These out-of equilibrium phenomena present new challenges for understanding the intricate dynamics that govern them. Our formulation within the Keldysh formalism provides an ideal foundation to explore the promising capabilities of two-dimensional spectroscopy in the study of nonequilibrium systems.
Finally, we see potential in using our approach to investigate disordered magnetic systems and address questions such as the formation of spin glasses, spin liquids, and fractionalized excitations, extending beyond the integrable limit, where most previous works have focused~\cite{wan2019resolving,Choi2020SpinLiquid,parameswaran2020asymptotically1,Spinon_Hart_2023,Fava2023divergent,fava2024longtimedivergencesnonlinearresponse}.

\vspace{0.5cm}

\section*{Acknowledgements}
We thank R. Andrei, L. Benfatto, U. Bhattacharya, J. Curtis, J. Fiore, A. Malyshev, F. Marijanovic, N. Sellati, and E. Vlasiuk for insightful discussions. A.G.S., I.M., and E.D. acknowledge the Swiss National Science Foundation (project $200021\_212899$) and ETH-C-06 21-2 equilibrium Grant with project number 1-008831-001 for funding. ED acknowledges support from the ARO grant number W911NF-21-1-0184. A.L. was supported by the U.S. Department of Energy, Office of Basic Energy Sciences, under Contract No. DE-SC0012704.


\appendix

\section{Perturbative calculation of \texorpdfstring{$\chi^{(3)}$}{x(3)} to lowest order in \texorpdfstring{$g$}{g}}\label{Appendix: tree_level}
As a first application of the laid groundwork in Sec.~\ref{sec: keldysh}, we show here an explicit calculation of $\chi^{(3)}$ at first-order in the interaction strength $g$, i.e., at the tree level:
\begin{widetext}
\begin{align}
    \tilde{\chi}^{(3)}_{\rm mf}(t_3,t_2,t_1)&=-8i\bigg{\langle}\varphi^{cl}(t_1+t_2+t_3)\varphi^{q}(t_1+t_2)\varphi^{q}(t_1)\varphi^{q}(0) \notag \\ &\times(-i)\int_{-\infty}^{\infty}dt'\int d^d \boldsymbol{r}'\,\frac{g}{3}\left[ \varphi^{cl}(t',r')\big{(}\varphi^q(t',r')\big{)}^3 + \big{(}\varphi^{cl}(t',r')\big{)}^3\varphi^q(t',r') \right]\bigg{\rangle} \notag \\ 
    & = -g\frac{8}{3}3!\int_{-\infty}^{\infty}dt'\, i\mathcal{D}^R_0(t_1+t_2+t_3-t')\, i\mathcal{D}^R_0(t'-t_1-t_2)\,i\mathcal{D}^R_0(t'-t_1)\,i\mathcal{D}^R_0(t') \notag \\
    & = -g\int_{-\infty}^{\infty}dt'\, \chi^{(1)}(t_1+t_2+t_3-t')\chi^{(1)}(t'-t_1-t_2)\chi^{(1)}(t'-t_1)\chi^{(1)}(t')
\end{align}
and expressed in frequency domain:
\begin{align}
    \tilde{\chi}^{(3)}_{\rm mf}(\omega_1,\omega_2,\omega_3) &\equiv \int dt_1\int dt_2\int dt_3 \,e^{i\omega_1t_1}e^{i\omega_2t_2}e^{i\omega_3t_3}\chi^{(3)}(t_1,t_2,t_3) = -g\, \chi^{(1)}(\omega_1)\chi^{(1)}(\omega_2-\omega_1)\chi^{(1)}(\omega_3-\omega_2)\chi^{(1)}(\omega_3). \label{eqn: chi_mf}
\end{align}
\end{widetext}
Performing the shift in the frequency arguments prescribed in equation~\eqref{eqn: phi3}, $\omega_1\rightarrow\omega_1$, $\omega_2\rightarrow\omega_1+\omega_2$, and $\omega_3\rightarrow\omega_1+\omega_2+\omega_3$ we arrive at
\begin{multline}
    \chi^{(3)}_{\rm mf}(\omega_1,\omega_2,\omega_3)=\tilde{\chi}^{(3)}_{\rm mf}(\omega_1+\omega_2+\omega_3,\omega_1+\omega_2,\omega_1)\\=-g\,\chi^{(1)}(\omega_1)\chi^{(1)}(\omega_2)\chi^{(1)}(\omega_3)\chi^{(1)}(\bar{\omega}),
\end{multline}
which provides $\chi^{(3)}$ to the lowest order in $g$, see Eq.~\eqref{eqn: mean_field_chi_3} in the main text.

\section{General Structure of \texorpdfstring{$\chi^{(3)}$}{x(3)}}\label{Appendix: effective_action}
In this Appendix we provide a systematic field-theoretical reformulation of the computation of $\chi^{(3)}$ in terms of the effective action. Even though this reformulation is general for arbitrary momentum coupling and measurement, we present it only for the zero momentum scenario in Eq.~\eqref{eqn: chi_3_decomposition}, which constitutes the central result of this Appendix. At this stage, we redefine for convenience a new source $J^{\alpha}(\boldsymbol{r},t) = 2j^\alpha(\boldsymbol{r},t)$, introduce the lumped space-time notation $x=(\boldsymbol{r},t)$, and define the one particle irreducible (1PI) effective action as the Legendre transform of $W[J] = i\log Z[J]$, such that:
\begin{gather}
    \Xi[\varphicl,\varphiq] = -W[\Jcl,\Jq] - \int dx \sum_\alpha J^{\bar{\alpha}}(x)\phi^\alpha(x), \\
     \frac{\delta W[J]}{\delta J^{\bar{\alpha}}(x)}= -\expval{\varphi^\alpha(x;J)} = -\phi^\alpha(x;J), \notag \\ \text{and} \quad  \frac{\delta\Xi[\phi]}{\delta\phi^\alpha(x)} = -J^{\bar{\alpha}}(x).
\end{gather}
Here, we have also introduced the convenient notations $\{\bar{\rm cl},\bar{\rm q}\}=\{\rm q,\rm cl\}$, and $\int dx = \int d^d \boldsymbol{r}\int dt$. The effective action fulfils the fundamental relation~\cite{dupuis2023field}
\begin{multline}
    \int dx'\sum_\gamma \frac{\delta^2W[J]}{\delta J^{\bar{\alpha}}(x_1)\delta J^{\bar{\gamma}}(x')}\frac{\delta^2\Xi[\phi]}{\delta \phi^\gamma(x')\delta \phi^\beta(x_2)}\\=\delta_{\bar{\alpha}\bar{\beta}}\,\delta(x_1-x_2). \label{eqn: fundamental_1PI}
\end{multline}
In particular, this implies that
\begin{gather}
    \frac{\delta^2\Xi[\phi]}{\delta\phi^\alpha(x_1)\delta\phi^\beta(x_2)} = \left(\mathcal{D}^{-1}\right)^{\alpha\beta}(x_1-x_2), \notag \\ \text{since} \quad \frac{\delta^2 W[J]}{\delta J^{\bar{\alpha}}(x_1)\delta J^{\bar{\beta}}(x_2)} = \mathcal{D}^{\alpha\beta}(x_1-x_2).
\end{gather}
The $n$th functional derivative of $\Xi$
\begin{gather}
    \Xi^{(n)}(\alpha_1x_1,\dots,\alpha_n x_n) = \frac{\delta^n\Xi[\phi]}{\delta\phi^{\alpha_n}(x_n)\dots\delta\phi^{\alpha_1}(x_1)}
\end{gather}
is often referred to as the $n$th 1PI vertex and can be connected to the $n$th-point connected Green's function. In particular, for the four-point  Green's function this relation is given by
\begin{widetext}
\begin{multline}
    \frac{\delta^4W[J]}{\delta J^{\bar{\alpha}_4}(x_4)\delta J^{\bar{\alpha}_3}(x_3) \delta J^{\bar{\alpha}_2}(x_2) \delta J^{\bar{\alpha}_1}(x_1)} = -\int dt'_{1} \int dt'_{2}\int dt'_{3}\int dt'_{4} \sum_{\alpha_1, \alpha_2, \alpha_3,\alpha_4} \Xi^{(4)}(\alpha_4'x_4',\alpha_3'x_3',\alpha_2'x_2',\alpha_1'x_1') \\  \times\mathcal{D}^{\alpha_4\alpha_4'}(x_4-x_4') \mathcal{D}^{\alpha_3\alpha_3'}(x_3-x_3') \mathcal{D}^{\alpha_1\alpha_2'}(x_2-x_2') \mathcal{D}^{\alpha_1\alpha_1'}(x_1-x_1').
    \label{eqn: W4-Gamma4}
\end{multline}
Here we have employed that $\Xi^{(3)}$ vanishes for the considered system. Hence, the third-order nonlinear response can be computed as
\begin{gather}
    \chi^{(3)}(x_3,x_2,x_1) = -8i\frac{\delta^4W[J]}{\delta \Jq(x_3+x_2+x_1)\delta \Jcl(x_2+x_1) \delta \Jcl(x_1) \delta \Jcl(0)}.
    \label{eqn: chi3_W}
\end{gather}
\end{widetext}
Assuming the system has translational invariance, equations~\eqref{eqn: W4-Gamma4} and~\eqref{eqn: chi3_W} can be combined to obtain the convenient expression
\begin{equation}
    \chi^{(3)}(p_3,p_2,p_1) = i\Gamma(p_3,p_2,p_1) \prod_i^4 \chi^{(1)}(p_i),
    \label{eqn: chi3_chi1_vertex_gen}
\end{equation}
where $p_i=(\boldsymbol{k}_i,\omega_i$), $p_4=(\bar{\boldsymbol{k}},\bar{\omega})$, and 
\begin{equation}
    \Gamma(p_3,p_2,p_1)=\frac{1}{2}\Gamma_{{\rm cl},{\rm q},{\rm q},{\rm q}}^{(4)}(\bar{p},-p_3,-p_2,-p_1).
\end{equation}
Note that in Eq.~\eqref{eqn: chi3_W} we have used that in the Keldysh formalism no vertex containing four external classical legs is possible; this would otherwise lead to a non-vanishing correction to the partition function~\cite{kamenev2023field}. For practical purposes when exciting a mode with light, we will be interested in coupling and measuring the zero momentum excitations of the system. Under these assumptions, we set all external momenta to zero and particularize Eq.~\eqref{eqn: chi3_chi1_vertex_gen} to arrive at 
\begin{equation}
    \chi^{(3)}(\omega_3,\omega_2,\omega_1) = i\Gamma(\omega_3,\omega_2,\omega_1) \prod_i^4 \chi^{(1)}(\omega_i) 
    \label{eqn: chi_3_decomposition},
\end{equation}
c.f. Eq.~\eqref{eqn: general_diag} of the main text.


\section{\texorpdfstring{$\chi^{(3)}$}{x(3)} within the lowest-order disorder approximation}
\label{Sec: tree_level_disorder}
\begin{figure}
    \centering
    \includegraphics[width=1\linewidth]{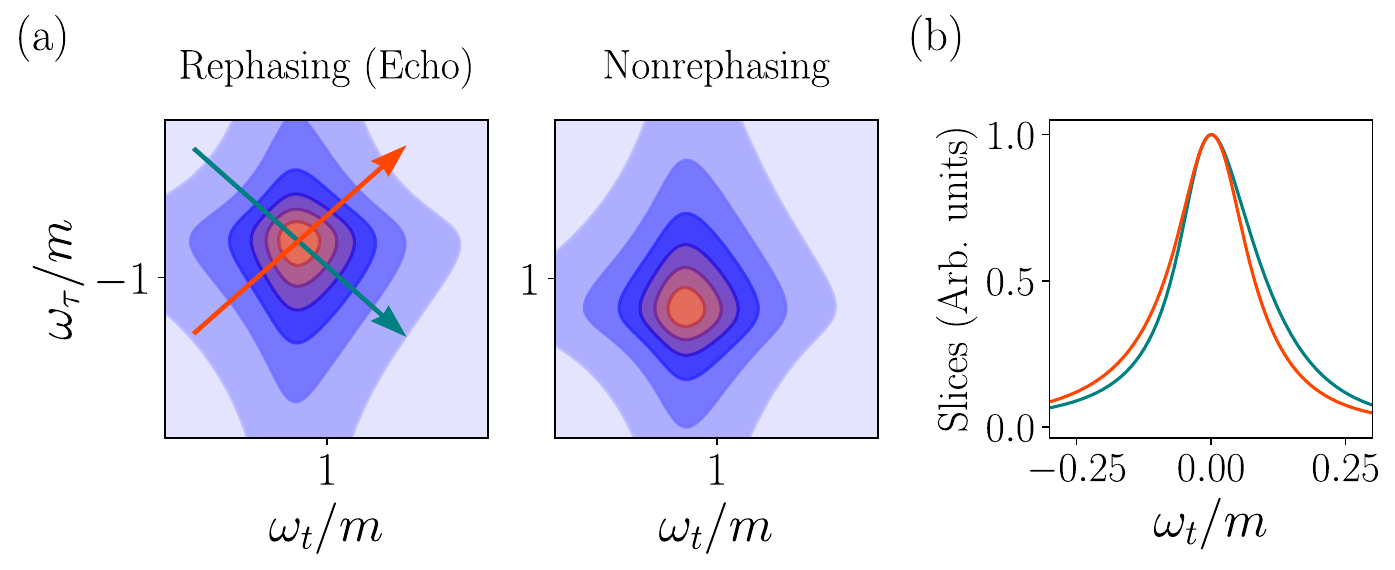}
    \caption{(a) Echo and nonrephasing peaks in the lowest-order disorder calculation with $V_0^2/m^2 = 2$, $\gamma/m=0.05$, and $g/m^2=0.01$. (b) Slices along the diagonal (teal) and cross-diagonal (orange) of the echo peak. As evident from the peaks itself and the cuts, this approximation is insufficient to capture and characterize the echo peak.}
    \label{fig: echo_lowest_order}
\end{figure}
In this Appendix, we briefly discuss the caclulation of $\chi^{(3)}$ in the lowest order approximation for the disorder. The employed Luttinger-Ward has the same structure as the one presented in Fig.~\ref{fig: LW_disorder} and is obtained by setting $\Gamma_{\rm dis}$ to be $\Gamma^{0}_{\rm dis}$, introduced in Eq.~\eqref{eqn: bare_disorder_vertex}. In similar fashion to the discussion of the main text, we compute the self-energy:
\begin{multline} 
    \Sigma^{R\,(0)}_{g-V}(\omega,\boldsymbol{k})=\vcenter{\hbox{\includegraphics[scale=0.75]{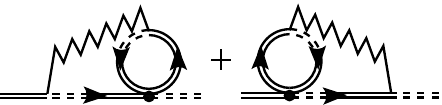}}}\\=2\times\frac{8igV_0^2}{L^{2d}}\sum_{\boldsymbol{k},\boldsymbol{q}}\int\frac{d\Omega}{2\pi}\,\mathcal{D}^R_{\boldsymbol{p}}(\omega)\mathcal{D}^R_{\boldsymbol{p}+\boldsymbol{q}-\boldsymbol{k}}(\Omega)\mathcal{D}^K_{\boldsymbol{q}}(\Omega),
\end{multline}
the vertex correction
\begin{align}
    \vcenter{\hbox{\includegraphics[scale=0.6]{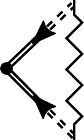}}} = &2igV_0^2\left(\frac{4}{L^d}\sum_{\boldsymbol{k}} \mathcal{D}_{\boldsymbol{k}}^R(\omega_a)\mathcal{D}_{-\boldsymbol{k}}^R(\omega_b)\right)\notag \\
    =&i2gV_0^2\lambda(\boldsymbol{0};\omega_a,\omega_b) \notag \\
    \vcenter{\hbox{\includegraphics[scale=0.65]{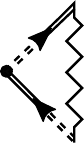}}}=&2igV_0^2\left(\frac{4}{L^d}\sum_{\boldsymbol{k}} \mathcal{D}_{\boldsymbol{k}}^R(\omega_a)\mathcal{D}_{\boldsymbol{k}}^R(\omega_b)\right) \notag \\
    = &2igV_0^2\lambda(\boldsymbol{0};\omega_a,\omega_b),
\end{align}
and the bubble in the presence of disorder:
\begin{multline} 
    g B_{\rm dis}^{(0)}(\omega)=\vcenter{\hbox{\includegraphics[scale=0.7]{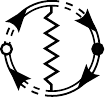}}}
    \notag \\ = 4igV_0^2\int \frac{d\Omega}{2\pi}\left(\frac{1}{L^d}\sum_{\boldsymbol{q}}\mathcal{D}_{\boldsymbol{q}}^{R}(\omega+\Omega)\mathcal{D}_{\boldsymbol{q}}^{K}(\Omega)\right)\\\times\lambda(\boldsymbol{0};\omega+\Omega,\Omega).
\end{multline}
This concludes the calculation of all the necessary ingredients to compute $\chi^{(3)}$ in the presence of disorder within our conserving approximation. $\chi^{(3)}$ is readily obtained following Eq.~\eqref{eq: vertex_elastic}. We proceed to compute the nonlinear maps, and present the rephasing and nonrephasing peaks in Fig.~\ref{fig: echo_lowest_order}. The echo peak shown in Fig.~\ref{fig: echo_lowest_order}(a) exhibits a slight asymmetry, which becomes more apparent when compared to the nonrephasing peak or investigating the slices in Fig.~\ref{fig: echo_lowest_order}(b). This minor asymmetry suggests that the calculation points in the correct direction, but is insufficient to capture the echo peak signature. A more sophisticated treatment, as the one presented in Sec.~\ref{sec: elastic}, is needed. 

\section{Inelastic Scattering}\label{sec: inelastic}
\begin{figure}
    \centering
    \includegraphics[width=\linewidth]{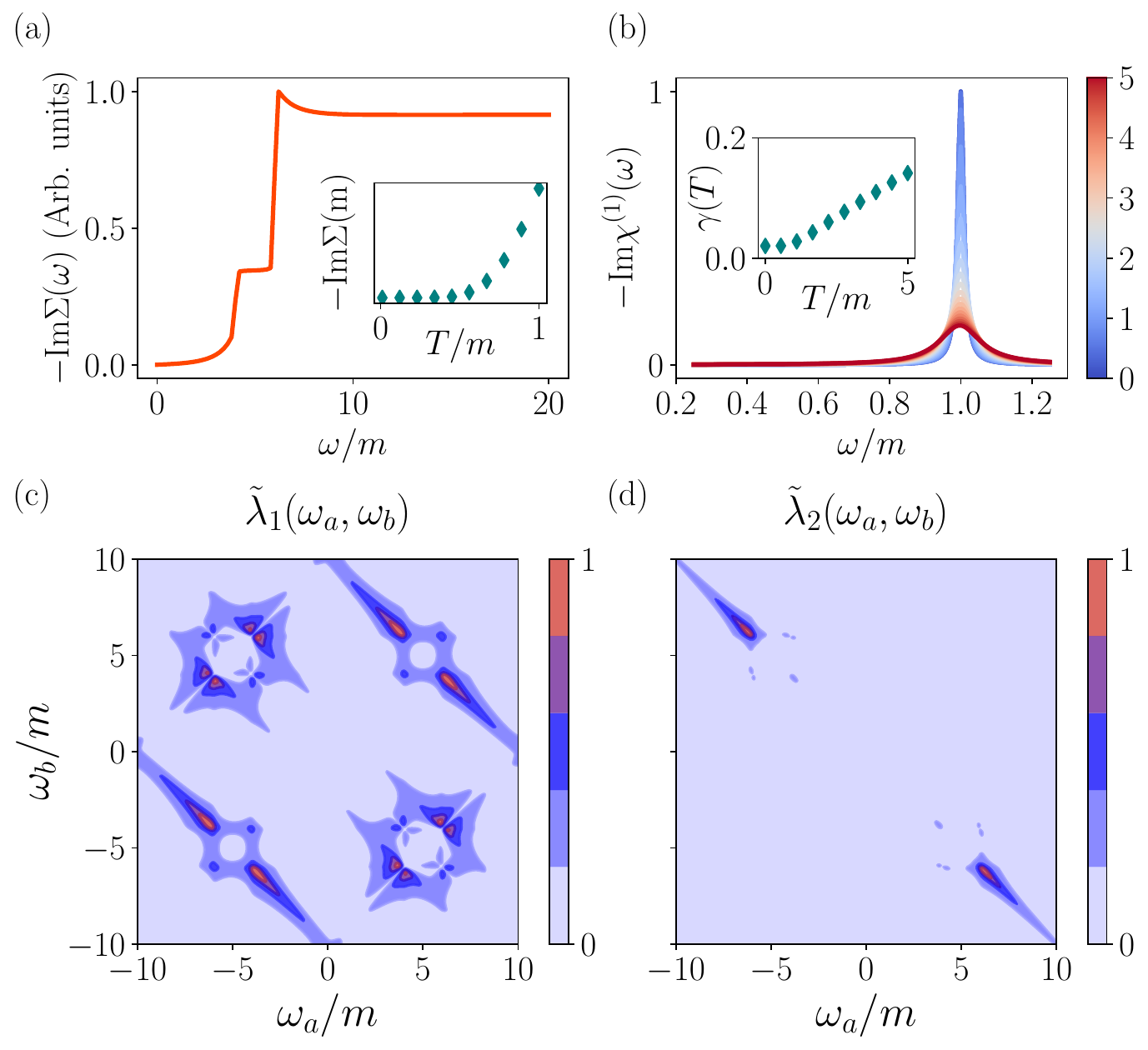}
    \caption{(a) Imaginary part of the self-energy due to the presence of the optical phonon, for $\mu/m^2=1$, $T/m=1$, and $M/m = 5$. In the inset, the on-shell broadening as a function of temperature. (b) Evolution of the boson linewidth due to the presence of the optical phonon as temperature increases. The inset provides the extracted values of $\gamma$ after a phenomenological fit to the self-energy. (c)-(d) Color-plot of the absolute value of the tree-level vertex correction for the two distinct vertices (Eqs.~\eqref{eqn: vertex_phonon_1} and~\eqref{eqn: vertex_phonon_2}) in the presence of the optical phonon; same numerical values as in (a).}
    \label{fig: phonon_supplement}
\end{figure}
In this Appendix, we present an analogous discussion to Sec.~\ref{sec: elastic} for the case of inelastic scattering. Diagrammatically, the analysis requires extra effort because an additional field must be incorporated to model the inelastic scattering events, introducing an additional energy scale, the mass of the mode $M$. In situations where $M/m\sim 1$, the modes can strongly couple and develop hybridized excitations, leading to a more complicated structure in the 2D map, as expected from the intuition of MDCS~\cite{hamm_zanni_2011,edler2003FermiResonance}.
The signatures of hybridization for collective modes in the many-body scenario require careful analysis and are therefore left for future work.

Here, we restrict ourselves to the scenario where the extra mode is an optical phonon with a mass much larger than the one of the collective mode under study, i.e. $M/m\gg 1$. 
In this regime, the phonon acts as an incoherent bath in which the collective mode can decay, leading to an $i\omega\gamma$ term in its self-energy, where $\gamma$ characterizes the broadening strength. 
Furthermore, the phonon can also, in principle, lead to vertex corrections in the four-wave mixing process. 
Intuitively, the rephasing of the inelastic scattering seems hard to conceive, since the characteristic frequency of the exchanged phonon is $\sim M$, much faster than the typical scale of the boson $m$. However, as a sanity check, we evaluate the vertex corrections coming from the phonon to confirm that no rephasing is possible. Thus, showing that these corrections can be completely disregarded in the regime $M/m\gg1$.

The interaction between the collective mode $\varphi$ and the optical phonon is given by
\begin{multline}
    S_{ph} = -\frac{\mu}{2}\int_{\mathcal{C}}dt\int d^d \boldsymbol{r}\, \psi(\boldsymbol{r},t)\varphi^2(\boldsymbol{r},t) \\= -\mu\int dt\int d^d \boldsymbol{r}\big{(}\varphi^{\rm cl}\varphi^{\rm cl}\psi^{q}+2\,\varphi^{\rm cl}\varphi^{\rm q}\psi^{\rm cl}+\varphi^{\rm q}\varphi^{\rm q}\psi^{\rm q}\big{)},
    \label{eqn: phonon-boson action}
\end{multline}
where we introduce classical and quantum fields for the phonon analogously to those for the collective mode.
Note that, although the coupling structure is similar to that of Eq.~\eqref{eqn: S_dis}, key differences arise due to the phonon's dynamics. First, the phonon enables energy transfer to the collective mode. Second, a larger number of diagrams must be considered in both the self-energy and vertex corrections.
\subsection{Self-energy}
The lowest order correction to the boson self-energy due to the presence of the optical phonon is given by
\begin{multline}
    \Sigma^R(\omega) = \vcenter{\hbox{\includegraphics[scale=0.7]{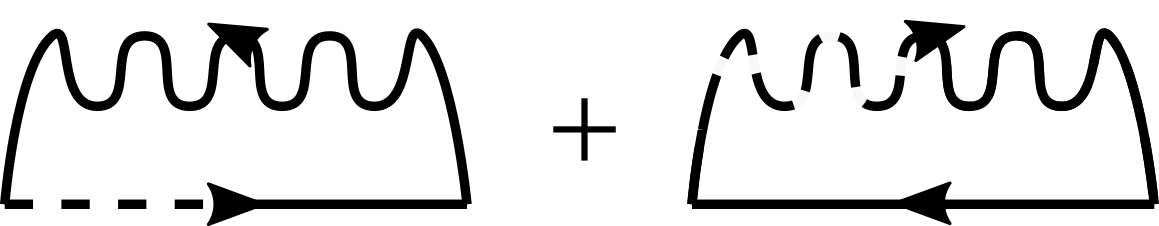}}} = \\ \frac{4\mu}{L^d}\sum_{\boldsymbol{k}} \left(\mathcal{D}^R_{0\boldsymbol{k}}(\omega) i\tilde{\mathcal{D}}^K_{0\boldsymbol{k}}(\omega) + i\mathcal{D}^K_{0\boldsymbol{k}}(\omega) \tilde{\mathcal{D}}^R_{0\boldsymbol{k}}(\omega) \right) 
\end{multline}
where $2\tilde{\mathcal{D}}^R_{\boldsymbol{k}}(\omega)=((\omega+i\eta)^2-M^2)^{-1}$ is the phonon retarded Green's function at momentum $\boldsymbol{k}$ and frequency $\omega$. 
To keep the mass of the excitation fixed at the physical mass $m$, we introduce a mass counterterm $\delta_m =\text{Re}\{\Sigma(\omega=m)\}$, such that $m^2=m_0^2+\Re \Sigma(m)-\delta_m$. 
Since the phonon can now exchange energy with the excitation of interest, the fluctuations of the phonon field control the linewidth broadening. At small temperatures, the phonon fluctuations are exponentially suppressed, and the broadening is negligible. Upon increasing temperature, the phonon fluctuations become sizable, resulting in a broadening of the boson linewidth, as can be seen in Fig.~\ref{fig: phonon_supplement} (a) and (b). This is a generic feature that can be phenomenologically captured, independently of the coupling strength $\mu$ or mass $M$, by a self-energy term of the form $i\omega\gamma(T)$, where $\gamma$ characterizes the broadening of the linewidth.

\subsection{Vertices}
Similarly to the case of elastic scattering, we evaluate the vertex corrections in the presence of the phonon.
Here, we compute the vertices only to lowest order to demonstrate that, in the limit $M/m\gg 1$, they can be completely disregarded. 
Unlike for the case of elastic scattering, Eq.~\eqref{eqn: elastic_vertex_correction}, the vertices cannot be reexpressed as a single vertex and have to be computed independently. 
These correspond to the exchange of a phonon between two incoming, or one incoming and one outgoing bosons and are given by
\begin{widetext}
\begin{align}
    \vcenter{\hbox{\includegraphics[scale=0.7]{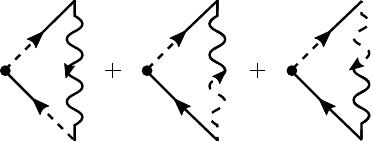}}}&=ig\tilde{\lambda}_1(\omega_a,\omega_b) = \frac{i8g\mu^2}{L^d}\sum_{0\boldsymbol{k}}\int\frac{d\Omega}{2\pi} \bigg{[} i\tilde{\mathcal{D}}_{0\boldsymbol{k}}^K(\Omega)\mathcal{D}_{0\boldsymbol{k}}^R(\omega_a+\Omega)\mathcal{D}_{0\boldsymbol{k}}^R(\omega_b+\Omega) \notag \\ 
    &+\tilde{\mathcal{D}}_{0-\boldsymbol{k}}^R(-\Omega)\mathcal{D}_{0\boldsymbol{k}}^R(\omega_a+\Omega)i\mathcal{D}_{0\boldsymbol{k}}^K(\omega_b+\Omega) +\tilde{\mathcal{D}}_{0\boldsymbol{k}}^R(\Omega)\mathcal{D}_{0\boldsymbol{k}}^R(\omega_a+\Omega)i\mathcal{D}_{0\boldsymbol{k}}^K(\omega_b+\Omega)\bigg{]}+ \omega_a\leftrightarrow\omega_b
    \label{eqn: vertex_phonon_1}
\end{align}
and
\begin{align}
    \vcenter{\hbox{\includegraphics[scale=0.7]{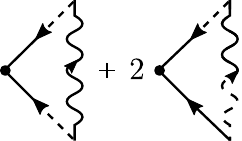}}}&=ig\tilde{\lambda}_2(\omega_a,\omega_b) =\frac{i8g\mu^2}{L^d}\sum_{\boldsymbol{k}}\int\frac{d\Omega}{2\pi} \bigg{[} i\tilde{\mathcal{D}}_{0\boldsymbol{k}}^K(\Omega)\mathcal{D}_{0\boldsymbol{k}}^R(\omega_a+\Omega)\mathcal{D}_{0-\boldsymbol{k}}^R(\omega_b-\Omega) \notag \\ 
    &+2\tilde{\mathcal{D}}_{0\boldsymbol{k}}^R(\Omega)\mathcal{D}_{0\boldsymbol{k}}^R(\omega_a+\Omega)i\mathcal{D}_{0-\boldsymbol{k}}^K(\omega_b-\Omega)\bigg{]}+ \omega_a\leftrightarrow\omega_b
    \label{eqn: vertex_phonon_2}
\end{align}
\end{widetext}
To show that there is no rephasing physics involved in these vertex corrections for $M/m\gg 1$, we present a color-plot of the absolute value of the two vertices $\tilde{\lambda}_1$ and $\tilde{\lambda}_2$ in Fig.~\ref{fig: phonon_supplement}(c) and (d) for $M/m=5$. 
Indeed, the fast dynamics of the phonon $\sim M$ push the effects of the vertex corrections to $\omega_a\sim\omega_b\sim M$, resulting in a negligible contribution in the frequency region of the boson mass $m$. 
However, as aforementioned, for $M/m\sim1$ the vertex corrections can be important around the mass energy $m$ and may give rise to interesting signatures in the 2D map.

As a last comment, we note that from the phonon-boson action in Eq.~\eqref{eqn: phonon-boson action} an effective vertex with external legs $cl-cl-cl-q$ arises, which contributes to the nonlinear response. However, the effective interaction scales as $\sim \mu^2/M^2$, for a large phononic mass.  Therefore, we can safely ignore its corrections to the preexisting interaction strength $g$.

\newpage

%


\end{document}